\documentclass[prc,aps,showpacs,showkeys,floatfix,groupedaddress,amsmath,amssymb]{revtex4}
\usepackage{graphicx} 
\usepackage{epsfig} 
\usepackage{dcolumn}
\usepackage{bm}

\newcommand{\q}{{\rm q}}
\newcommand{\p}{{\rm p}}
\newcommand{\rmRe}{{\rm Re}\:}
\newcommand{\qv}{\mathbf{q}}
\newcommand{\rv}{\mathbf{r}}
\newcommand{\uv}{\mathbf{u}}
\newcommand{\kv}{\mathbf{k}}
\newcommand{\kpv}{\mathbf{k}^{\prime}}
\newcommand{\mH}{\mathcal H}
\newcommand{\mE}{\mathcal E}
\newcommand{\be}{\begin{eqnarray}}
\newcommand{\ee}{\end{eqnarray}}

\begin{document}

\title{Cluster formation in asymmetric nuclear matter:\\
semi-classical and quantal approaches}

\author{C. Ducoin} 
\affiliation
{Istituto Nazionale di Fisica Nucleare, Sezione di Catania, Via Santa Sofia 64, I-95123 Catania, Italy\\
LPC (IN2P3-CNRS/Ensicaen et Universit\'{e}), F-14050 Caen CEDEX, France\\
GANIL (DSM-CEA/IN2P3-CNRS), B.P.5027, F-14076 Caen CEDEX 5, France}  
\author{J. Margueron}
\affiliation{Institut de Physique Nucl\'eaire, IN2P3-CNRS and
Universit\'e Paris-Sud, F-91406 Orsay Cedex, France\\
Center for Mathematical Sciences, University of Aizu, Aizu-Wakamatsu, 965-8580 Fukushima, Japan}
\author{Ph. Chomaz} 
\affiliation{GANIL (DSM-CEA/IN2P3-CNRS), B.P.5027, F-14076 Caen CEDEX 5, France} 



\begin{abstract}
The nuclear-matter liquid-gas phase transition induces instabilities against finite-size density fluctuations.
This has implications for both heavy-ion-collision and compact-star physics.  
In this paper, we study the clusterization properties of nuclear matter
in a scenario of spinodal decomposition,
comparing three different approaches:
the quantal RPA, its semi-classical limit (Vlasov method), and a hydrodynamical framework.
The predictions related to clusterization are qualitatively in good agreement 
varying the approach and the nuclear interaction.
Nevertheless, it is shown that 
i) the quantum effects reduce the instability zone, and disfavor short-wavelength fluctuations;
ii) large differences appear between the two semi-classical approaches,
which correspond respectively to a collisionless (Vlasov) and local equilibrium description (hydrodynamics);
iii) the isospin-distillation effect is stronger in the local equilibrium framework;
iv) important variations between the predicted time-scales of cluster formation appear
near the borders of the instability region.
\end{abstract}

\maketitle

\section{Introduction}
\label{SEC:introduction}

In infinite nuclear matter, 
a phase transition of the liquid-gas type~\cite{Bertsch-PLB126,Finn-PRL49} is predicted to
occur at sub-saturation density with a critical temperature of about 15~MeV, 
depending on the theoretical models.
This phenomenon corresponds to the separation of the system 
into two macroscopic (infinite) phases of different densities.
This thermodynamic property is at the origin of finite-size spinodal instabilities,
leading to the clusterization of the homogeneous medium.
In this process, in contrast with the thermodynamic case,
the Coulomb interaction can be included and density-gradient terms come into play:
the liquid-gas instabilities can then be related to the formation of fragments
occurring in finite nuclear systems.
In heavy-ion collisions performed in nuclear facilities, around the Fermi energy,
several signals of the liquid-gas phase transition have been analyzed~\cite{Das-PhysRep}.
In particular, the spinodal instabilities could be at the origin of the
multifragmentation process~\cite{Borderie-NPA734,Rivet-NPA749,Trautmann-NPA-752}.
Several models have been built in order to describe the dynamics 
leading to the formation of clusters~\cite{PC-PhysRep,Baran-PhysRep410}.
Clustering is also expected to play an important
role in compact stars~\cite{Glendenning-PhysRep,Lattimer-PhysRep}.  
It is predicted to lead to the presence of complex structures
like the "pasta phases" occurring in the inner crust of neutron 
stars~\cite{Ravenhall-PRL50,Lassaut-AA183,Watanabe-PRC68,Horowitz-PRC69}.
At lower densities and finite temperature, 
star matter may clusterize into light nuclei~\cite{hor06,bey00}.
Around the core of supernovae, 
density fluctuations could have important effects on neutrino propagation~\cite{Margueron-PRC70}, 
which is crucial for the description of supernova dynamics~\cite{Buras-PRL90}.

This paper addresses the clusterization properties of an infinite nuclear medium:
the onset of the liquid-gas instabilities is characterized 
by studying the incidence of a small density fluctuation introduced in the homogeneous medium.
Such approach is based on a spinodal-decomposition scenario,
which assumes that fragments are formed in an out-of-equilibrium process.
Indeed, if a homogeneous system is quenched into the region of finite-size instabilities, 
it will decompose into clusters:
in the spinodal-decomposition scenario, the early dynamics governing this evolution
consists in the rapid amplification of the unstable fluctuations,
whose properties can be understood in the small-amplitude limit.
For a fast reaction, fragments could be separated before equilibrium is realized,
thus keeping the memory of the bulk instability.
Identifying fragments produced through spinodal decomposition is then of great importance
to obtain informations about the homogeneous-matter energetics at sub-saturation density.
Experimental indications of this mechanism in heavy-ion collisions
have been obtained 
for instance through the analysis of size correlations~\cite{Borderie-PRL86,Tabacaru-EPJA18}.
We should note that the study of finite-size instabilities in homogeneous matter
is complementary to the description of an equilibrated, clusterized system,
as can be given at very low density by the virial equation of state~\cite{hor06}.
With this respect, the finite-size instability region we discuss can be viewed as the minimal region
where the equilibrated system is formed of clusters.

The model of infinite nuclear matter provides a useful insight into the behavior of nuclear systems,
revealing robust qualitative features whose origin can be clearly analysed.
The spinodal decomposition studied in this framework
can be related to the nuclear multifragmentation process
by neglecting the finite size of the homogeneous medium formed in the early stage of the collision.
It also gives a picture of the instability against clusterization in compact-star matter,
where the degree of freedom associated with the electrons can be neglected~\cite{CD-A3}.

Our calculations are based on Skyrme density functionals.
Specific attention is paid to the isospin degree of freedom,
relevant for the physics of compact stars and exotic nuclei.
The study of the instabilities is treated in three different approaches.
Starting from the thermodynamic definition of the liquid-gas phase transition,
we first concentrate on the free-energy curvature properties.
This leads us to study the finite-size density fluctuations 
in a Thomas-Fermi approach~\cite{CD-A3,Pethick-NPA584},
defining a static criterion for the existence of instabilities based on the free-energy variations.
Afterward, consistently with this local equilibrium framework, 
we treat the time evolution by introducing hydrodynamical equations.
This defines a first semi-classical approach of the clusterization dynamics.
The second approach is fully quantal and dynamical:
it is based on the linearization of the Time-Dependent Hartree-Fock theory 
(TDHF)~\cite{Ring-Schuck}.
This corresponds to the random-phase approximation (RPA).
In this case, the evolution of the instability is described in the collisionless limit.
The differences expected between hydrodynamics and RPA results have then two origins:
the quantal effects and the collisionless description.
In order to disentangle these two effects, we will also consider a third approach,
the Vlasov method which is the semi-classical limit of the RPA.
It then includes the collisionless description but not the quantal effects.

Different Skyrme forces have been used in order to check the
dependence of the results on the choice of a specific set of parameters.  
We take as a reference the Skyrme-Lyon parameterization SLy230a~\cite{Chabanat-NPA627}, 
a modern force designed to treat neutron-rich matter,
already employed for astrophysical applications~\cite{Douchin-PLB485, Douchin-NPA665}.  
Two other Skyrme forces are used for comparison: 
RATP~\cite{Rayet-AA116} - the first interaction which includes neutron
matter into the fitting procedure - and SGII~\cite{Giai-NPA371},
a typical widely used parameterization.
The interaction RATP is constrained by the neutron-matter equation of state 
calculated by Friedman and Pandharipande in the 80ies~\cite{FriedPandha-NPA361}, 
while the interaction SLy230a uses a more recent neutron-matter equation of state
based on an improved realistic interaction~\cite{Wiringa-PRC38}.
Let us remark that these interactions 
have quadratic velocity-dependent terms which modify the single-particle energies 
and give rise to effective masses $\mathrm{m}^*_i$ ($i$=n, p).
The effects of an isovector momentum-dependent mean field
on nuclear-matter properties have been studied in Ref.~\cite{Xu-PRC77}.

\begin{figure}[t]
\begin{center}
\includegraphics[width=0.4\linewidth]{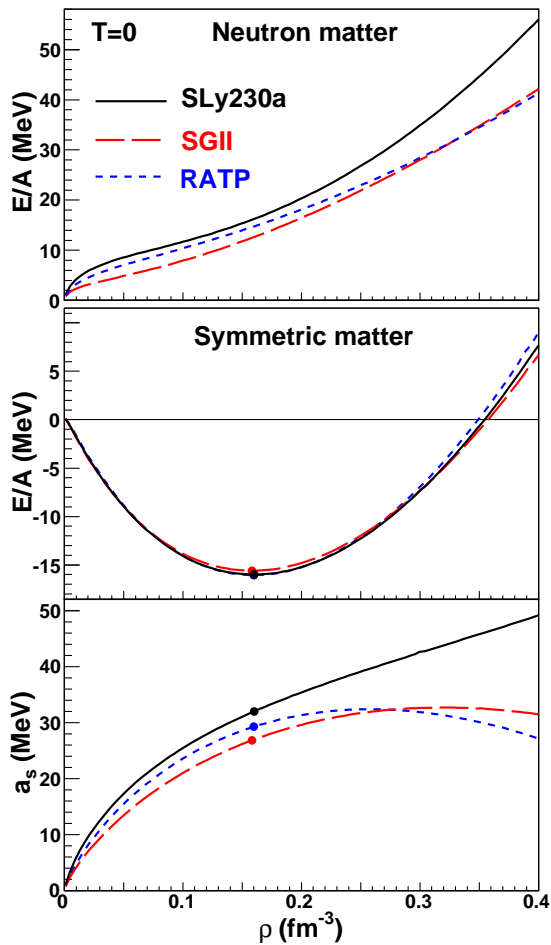}
\end{center}
\caption{(color online) Binding energy in pure neutron matter (top),
	in symmetric 
	nuclear matter (middle) and symmetry energy (bottom) versus
	the density for the three Skyrme interactions considered
	in this paper (SLy230a, SGII and RATP). The saturation point
	in symmetric nuclear matter is also indicated (dots).}
\label{FIG:carac_3F}
\end{figure}

For these three interactions, Fig.~\ref{FIG:carac_3F} shows as a function of the density
the binding energy $\mathrm{E/A}$ in symmetric nuclear matter and in pure
neutron matter, as well as  the symmetry energy $\mathrm{a}_{\rm s}$. 
The different saturation properties are reported in Tab.~\ref{tab:satur}.
One can see that the predictions of scalar properties, saturation density, energy and incompressibility
are very close for these realistic forces.
The values of the symmetry energy at saturation are also quite close,
the main difference being its density dependence, as shown on Fig.~\ref{FIG:carac_3F}.
The symmetry energy slope $\mathrm{L}$ and curvature $\mathrm{K}_{\rm sym}$ 
are also indicated in Tab.~\ref{tab:satur}.
It should be noticed that the selected Skyrme forces present an "asy-soft" behaviour, 
namely values of $\mathrm{L}$ lower than 60~MeV, while
recent constraints extracted from different experimental analysis indicate 
that $\mathrm{L}$ shall be larger than about
60~MeV~\cite{Chen-PRC72,Piekarewicz-PRL95,Shetty-PRC75}.
This however has no 
drastic effects on the results we present, since the spinodal contour in
asymmetric nuclear matter is non linearly related to the symmetry
energy and its first and second derivatives, \emph{cf} Eq.~(\ref{eq:Ctilde}).
A systematic analysis of the spinodal properties for various
relativistic and non-relativistic interactions, "asy-soft" and
"asy-stiff", is in preparation~\cite{Skyrme-RMF-prepa}.

\begin{table}[t]
\begin{center}
\renewcommand{\arraystretch}{1.5}
\begin{tabular}{|c|c|c|c|c|c|c|c|}
\hline 
 & $\rho_{0}$ & $\mathrm{E/A}$ & $\mathrm{K}_{\infty}$ & $\mathrm{a}_{{\rm s},0}$ & $\mathrm{L}$ & $\mathrm{K}_{\rm sym} $ & $T_c$ \\[-0.15cm]
 & \; [fm$^{-3}$] \;& \;[MeV] \;& \;[MeV]\; & \;[MeV]\; & \;[MeV]\; & \;[MeV]\; & \;[MeV]\; \\
\hline 
\;SLy230a\; & $0.160$ & $-15.99$ & $229.9$ & $31.98$ & $44.3$ & $-98.2$ &14.55 \\
\hline 
SGII & $0.158$ & $-15.59$ & $214.7$ & $26.83$ & $37.6$ & $-145.9$ & 14.46 \\
\hline 
RATP & $0.160$ & $-16.05$ & $239.6$ & $29.26$ & $32.4$ & $-191.3$ & 14.72 \\
\hline
\end{tabular}
\end{center}
\caption{Saturation properties of the three Skyrme forces SLy230a,
  SGII and RATP. Are given the values for the saturation density
  $\rho_0$: 
  the binding energy $\mathrm{E/A}$; 
  the incompressibility modulus $\mathrm{K}_\infty$; 
  the symmetry energy $\mathrm{a}_{{\rm s},0}$; 
  the symmetry energy slope $\mathrm{L}=3\rho_0(\partial \mathrm{a}_{\rm s}/\partial\rho)$;
  the symmetry energy curvature $\mathrm{K}_{{\rm sym}}=9\rho_0^2(\partial^2\mathrm{a}_{\rm s}/\partial\rho^2)$;
  the critical temperature of the liquid-gas phase transition $T_c$.}
\label{tab:satur}
\end{table}%

The paper is organized as follows.
Before discussing the finite-size instabilities, we address
the thermodynamic spinodal properties in Sect.~\ref{SEC:thermo}.
The hydrodynamical description for the clusterization process is treated in Sect.~\ref{SEC:local equilibrium}.
The collisionless dynamics based on the TDHF equation is treated in Sect.~\ref{SEC:collisionless}, 
where the quantal RPA and Vlasov methods are addressed. 
We discuss the different results in Sect.~\ref{SEC:discussion},
and conclusions are drawn in Sect.~\ref{SEC:conclusion}.

\section{Spinodal region from a thermodynamical approach}
\label{SEC:thermo}

In this section, we address the thermodynamic liquid-gas phase transition of nuclear matter,
where each phase is infinite and homogeneous.
This allows a geometrical representation of the phase equilibrium, 
obtained by a Gibbs construction on the free-energy surface:
thus can be emphasized the fundamental role of the free-energy curvature in the liquid-gas instabilities
studied in the following sections. 

The thermodynamic equilibrium of a system 
is given by maximizing the entropy in the space of observables.
For homogeneous nuclear matter, this surface presents 
an anomalous convexity in the phase-transition region.
In the thermodynamic limit where interface contribution is neglected,
this convexity can be corrected maximizing the entropy 
by linear interpolations which define pairs of phases in coexistence.
In other words, the thermodynamic equilibrium is given by the Gibbs construction, 
which delimits the phase-transition boundaries.
This coexistence region contains the spinodal region,
where homogeneous matter is locally unstable 
(the entropy surface being locally convex).
For a given value of the temperature, 
this is equivalent to a negative value of the free-energy curvature matrix,
which we now determine.

In the Skyrme density-functional theory (DFT), 
the average energy density of homogeneous nuclear matter 
$\mathcal H = \langle \hat{\mathrm{H}}\rangle/V$ is a function of the one-body density matrix.  
The full expression of $\mathcal H$ obtained from a Skyrme-like interaction
is given for example in Ref.~\cite{Chabanat-NPA627}.
Considering homogeneous, spin-saturated matter with no Coulomb interaction,
the energy-density functional reduces to:
\begin{eqnarray}
\label{EQ:E_HNM}
\mathcal{H} = \mathcal{H}^{\rm h}
&=& \frac{\hbar ^{2}}{2m}\tau 
+C_{\rm eff}\rho \tau+D_{\rm eff}\rho _{3}\tau _{3} \nonumber \\
&&+C_{0}\rho ^{2}+D_{0}\rho_{3}^{2} +C_{3}\rho ^{\sigma +2}
+D_{3}\rho^{\sigma }\rho _{3}^{2} 
\; ,
\end{eqnarray}
where the superscript "$\rm h$", introduced for later use, means "homogeneous".
Coefficients $C$ an $D$ are linear combinations of the
standard Skyrme parameters \cite{Chabanat-NPA627,CD-A1}.  
We have introduced the isoscalar and isovector particle densities, $\rho$ and $\rho_3$, 
as well as kinetic densities, $\tau$ and $\tau_3$: 
\begin{equation}
\begin{array}{lll}
\rho =\rho _{\rm n}+\rho _{\rm p} & ; & \tau =\tau _{\rm n}+\tau _{\rm p} \\ 
\rho _{3}=\rho _{\rm n}-\rho _{\rm p} & ; & \tau _{3}=\tau_{\rm n}-\tau _{\rm p}
\end{array}
\end{equation}
where, denoting $i$ the third component of the isospin
($\rm n$ for neutrons and $\rm p$ for protons),
the kinetic densities are defined by $\tau_i=\langle \hat{\p}^2\rangle_i/\hbar^2$.
The energy density $\mH$ is then a function of neutron and proton densities $\{\rho_i,\tau_i\}$.
From the Skyrme functional is derived the mean field for each type of
particle $i=(\rm n,\rm p)$. 
In uniform matter, 
the mean field reduces to a potential $\mathrm{U}_i$ and an effective mass $\mathrm{m}_i^*$
such that the single-particle energies read:
\be
\epsilon_i(p)=\frac{p^2}{2\mathrm{m}_i^*}+\mathrm{U}_i 
\; ; \;\;\;
\mathrm{U}_i=\frac{\partial\mH}{\partial\rho_i}
\; ; \;\;\;
\frac{\hbar^2}{2\mathrm{m}_i^*}=\frac{\partial\mH}{\partial\tau_i}
\; ,
\ee
where $p$ is the momentum quantum number of the free particles.
In the grand-canonical formalism, individual levels are occupied according
to the Fermi-Dirac statistics with occupation numbers 
$\mathrm{n}_i(p)=\left[1+\exp(\beta (\epsilon_i(p)-\mu_i)) \right]^{-1}$, 
where $\beta=1/T$ is the inverse temperature, 
and $\mu_i$ the chemical potential of the corresponding nucleon type $i$. 
The mean-field grand-canonical partition function $\mathrm{Z}_0$ 
is formally identical to the free-Fermi-gas one, 
replacing $\mu_i$ by $\nu_i=\mu_i-\mathrm{U}_i$ and $\mathrm{m}_i$ by $\mathrm{m}_i^*$.
At given values of $T$, $\rho_{\rm n}$ and $\rho_{\rm p}$, 
chemical potentials and kinetic densities are fixed, 
determining the energy density $\mH$ 
and the mean-field entropy density $\mathrm{s}$, directly related to $\mathrm{Z}_0$. 
The free-energy density $\mathrm{f}(T,\rho_{\rm n},\rho_{\rm p})$ is deduced
from the Legendre transform $\mathrm{f}=\mH-T\mathrm{s}$.
We denote $\mathrm{f}^{\rm h}$ the free-energy density of a homogeneous system:
\begin{equation}
\label{eq:fh}
\mathrm{f}^{\rm h}(T,\rho_{\rm n},\rho_{\rm p})=\mH^{\rm h}-T\mathrm{s}
\; .
\end{equation}
In the following, working at constant temperature, 
the dependence of the functions on $T$ will be implicit for a more concise notation.
For a given temperature, the thermodynamic spinodal region 
corresponds to a negative curvature of the surface 
$\mathrm{f}^{\rm h}(\rho_{\rm n},\rho_{\rm p})$,
\emph{i.e.} a negative eigen-value of the curvature matrix 
$\mathrm{C}^{\rm h}=\{C^{\rm h}_{ij}\}$, with:
\be
\mathrm{C}^{\rm h}_{ij} (\rho_{\rm n},\rho_{\rm p}) 
= \frac{ \partial^2 \mathrm{f}^{\rm h}}{ \partial\rho_j\partial\rho_i}
= \frac{\partial\mu_i}{ \partial\rho_j}
\; ,
\ee
whose lower eigen-value is denoted $\mathrm{C}^{\rm h}_<$.

\begin{figure}
\begin{center}
\includegraphics[width=0.5\linewidth]{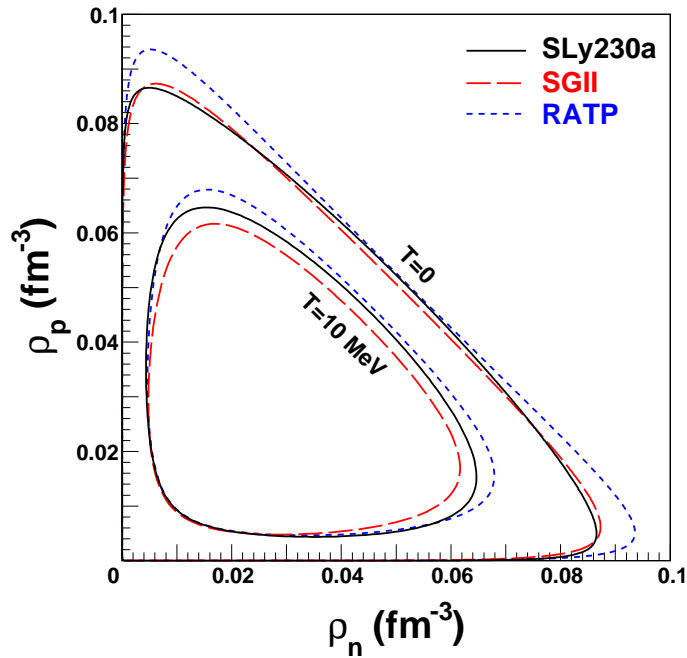}
\caption{(color online) Contours of the thermodynamic spinodal region 
for two values of the temperature ($T=0$ and $T=10$~MeV) 
and for the Skyrme-like interactions: SLy230a, SGII and RATP.}
\label{FIG:spino_rnrp_3F}
\end{center}
\end{figure}

Figure~\ref{FIG:spino_rnrp_3F} shows the spinodal contours
obtained for two values of temperature ($T=0$ and $10$~MeV)
and three Skyrme-like interactions: SLy230a, SGII and RATP.
Small quantitative differences can be noticed between these interactions, 
especially at large asymmetry and density.
These differences can be analyzed
in terms of isovector properties of the density functional at ($\rho_{\rm n}=\rho_{\rm p}$),
which we will now discuss. 
We introduce the variables $(\rho,y)$
where $\rho$ is the total density and $y=\rho_3/\rho$ the asymmetry.
Let us focus on the point $(\rho=\rho_{\rm s},y=0)$,
defined as the high-density border of the spinodal for symmetric matter:
the extension of the spinodal region in the asymmetry direction
depends on the shape (convexity) of the spinodal contour at this point.
This contour is defined by $\mathrm{C}^{\rm h}_<(\rho,y)=0$,
where $\mathrm{C}^{\rm h}_<$ is the lower eigen-value of the free-energy curvature matrix.
The isospin symmetry imposes 
$\left[ \partial \mathrm{C}^{\rm h}_</\partial y|_{\rho}\right](\rho,0)=0$.
The convexity of the spinodal contour at the point $(\rho_{\rm s},0)$ 
can then be characterized by the quantity
$\tilde{\mathrm{C}_{\rm s}} = \left[ \partial^2 \mathrm{C}^{\rm h}_</\partial y^2|_{\rho}\right] (\rho_{\rm s},0)$.
Indeed, if $\tilde{\mathrm{C}_{\rm s}}$ is positive, for a small asymmetry $y$ we have
$\mathrm{C}^{\rm h}_<(\rho_{\rm s},y)>0$, meaning that the point $(\rho_{\rm s},y)$ is outside the spinodal contour.
Thus, a positive (negative) value of $\tilde{\mathrm{C}_{\rm s}}$ 
corresponds to a concave (convexe) shape of the spinodal contour 
at point $(\rho=\rho_{\rm s},y=0)$.
The value of $\tilde{\mathrm{C}_{\rm s}}$ depends on the symmetry energy 
and its density derivatives:
\be
\label{eq:Ctilde}
\tilde{\mathrm{C}}_{\rm s} 
&=& \frac{4}{\rho \, \mathrm{a}_{\rm s}} \left[ \mathrm{a}_{\rm s} (\rho^2\mathrm{a}_{\rm s}'' +2\rho \mathrm{a}_{\rm s}') - 2 (\rho \mathrm{a}_{\rm s}^{\prime})^2
\right] 
\ee
where $\mathrm{a}_{\rm s}^{\prime}=\left[ \partial \mathrm{a}_{\rm s}/\partial \rho|_y\right](\rho=\rho_{\rm s},y=0)$
and $\mathrm{a}_{\rm s}^{\prime\prime}=\left[ \partial^2 \mathrm{a}_{\rm s}/\partial \rho^2|_y\right](\rho=\rho_{\rm s},y=0)$.
The respective contributions entering in Eq.~(\ref{eq:Ctilde})
are shown in Tab.~\ref{tab:Ctilde} for the three Skyrme-like interactions SLy230a, SGII and RATP. 
It is clear that none of these terms dominates the sum. 
Information on the density dependence of the symmetry energy
(slope and curvature with respect to $\rho$) 
is then determinant to predict the qualitative behaviour of the spinodal.
Conversely, information on the evolution of the spinodal extension with the asymmetry
would bring constraints on the relation between these terms according to Eq.~(\ref{eq:Ctilde}).

\begin{table}[t]
\begin{center}
\renewcommand{\arraystretch}{1.5}
\begin{tabular}{|c|c|c|c|c|c|}
\hline 
& $\rho_{\rm s}$ & $\mathrm{a}_{\rm s}$ & $\rho_{\rm s} \mathrm{a}_{\rm s}'$ & 
$\rho_{\rm s}^2 \mathrm{a}_{\rm s}''$ & $\tilde{\mathrm{C}}_{\rm s}$ \\ [-0.15cm]
& \;[fm$^{-3}$] \;& \;[MeV]\; & \;[MeV]\; & \;[MeV]\; & \;[MeV.fm$^3$]\; \\
\hline 
\;SLy230a\; & $0.1023$ & $25.73$ & $13.21$ & $-9.70$ & $123.2$\\
\hline 
SGII & $0.1006$ & $21.06$ & $12.41$ & $-9.33$ & $34.3$\\
\hline 
RATP & $0.1026$ & $23.88$ & $12.69$ & $-12.41$ & $-20.2$\\
\hline
\end{tabular}
\end{center}
\caption
{Contributions to the quantity $\tilde{\mathrm{C}}_{\rm s}$ coming from the symmetry
energy and its derivatives at point $\rho_{\rm s}$ 
(upper border of the spinodal in symmetric matter).
The value of  $\rho_{\rm s}$ is also indicated. 
The Skyrme-like interactions SLy230a, SGII and RATP have been considered.}
\label{tab:Ctilde}
\end{table}%

\section{local equilibrium description of cluster formation}
\label{SEC:local equilibrium}

Let us now turn to finite-size density fluctuations, 
relevant for the study of matter clusterization in neutron-star crust and supernova core, 
as well as nuclear multifragmentation.
Coulomb and surface energy now contribute to reduce 
the instability induced by the thermodynamic spinodal.
However, a region of finite-size instabilities can still be found,
causing the instability of homogeneous matter against cluster formation.
This region is determined considering plane-wave density fluctuations,
varying the corresponding wave-number $\qv$:
indeed, any density fluctuation can be expressed as a 
coherent sum of plane waves
and the instability with respect to at least one $\qv$ value
signs the presence of finite-size instabilities.

This section presents a first approach for the study of the unstable region:
we assume local equilibrium in a semi-classical framework,
the local densities defining local Fermi momenta.
First, an instability criterion is defined 
from a Thomas-Fermi calculation of the free-energy variation;
then hydrodynamical equations are introduced in order to describe the temporal evolution.

\subsection{Finite-size instabilities in the Thomas-Fermi approach}
\label{SSEC:Thomas-Fermi}

We detail here how the free-energy variation induced by a small density fluctuation
is derived from the Thomas-Fermi approximation. 
Let us consider a plane-wave density fluctuation of wave number $\qv$, 
affecting both the neutron and proton densities $(i={\rm n},{\rm p})$:
$\rho_i = \rho_{0,i} + \delta \rho_i$, with 
$\delta \rho_i = \mathrm{A}_i {\rm e}^{\rm i\qv \cdot \rv} + cc$.
The size of a density fluctuation, $\lambda$, is characterized by the
momentum $\q$: $\lambda=2\pi/\q$
(note that the thermodynamic limit is obtained back 
with the infinite-wavelength density fluctuation, namely $\q \rightarrow 0$).
The Thomas-Fermi approximation supposes that density variations 
$\delta \rho_i$ are smooth enough to allow 
the definition of a local Fermi momentum.
This approximation is valid if the number of states $\rho_i \Omega$ - where $\Omega$ is the volume - 
is very large for a typical scale of $\lambda$. 
Taking for the volume $\Omega=\lambda^3$,
this condition leads to the following criterium: 
$4\pi g(k_{\rm F_i}/\q)^3/3\gg 1$, 
where $g$ is the spin degeneracy $g=2$. 
It is then clear that this approach is valid if $\q$ is smaller than
the Fermi momentum $k_{\rm F_i}$~\cite{Ayik-PLB353,PC-PhysRep}.

In the description of nuclear matter with density fluctuations,
finite-size effects appear because of the momentum dependence of the nuclear force,
which is also related to its range.
In the Skyrme framework, the surface energy can be written as:
\begin{equation}
\label{eq:Hnabla}
\mathcal H^{\nabla}
=C_{\rm nn}^{\nabla}(\nabla \rho_{\rm n})^2
+C_{\rm pp}^{\nabla}(\nabla \rho_{\rm p})^2
+2C_{\rm np}^{\nabla} \: \:  \nabla \rho_{\rm n} \cdot \nabla \rho_{\rm p},
\end{equation}
where coefficients $C_{ij}^{\nabla}$ are combinations of the standard Skyrme parameters:
\begin{equation}
\begin{array}{l}
C_{\rm nn}^{\nabla}=C_{\rm pp}^{\nabla}
=\frac{3}{32} \left[ t_1(1-x_1) - t_2(1+x_2) \right] \;,\\
C_{\rm np}^{\nabla}=C_{\rm pn}^{\nabla}
=\frac{1}{32} \left[ 3t_1(2+x_1) - t_2(2+x_2) \right] \;.
\end{array}
\end{equation}
To get a deeper insight, 
we can also express $\mH^{\nabla}$ in terms of the isoscalar and isovector density gradients, namely:
\be
\mH^{\nabla}=C_{11}^{\nabla}(\nabla\rho)^2+C_{33}^{\nabla}(\nabla\rho_3)^2
\ee
where the isoscalar and isovector coefficients are related to (\ref{eq:Hnabla}) as:
\be
\begin{array}{l}
C_{\rm nn}^{\nabla}=C_{\rm pp}^{\nabla}=C_{11}^{\nabla}+C_{33}^{\nabla} \;,\\
C_{\rm np}^{\nabla}=C_{\rm pn}^{\nabla}=C_{11}^{\nabla}-C_{33}^{\nabla} \;.
\end{array}
\ee
The values of the coefficients $C^{\nabla}$ associated with the density-gradient dependence of the Skyrme functional
are given in Tab.~\ref{TAB:coef-Hfin} for several interactions.
We can notice that the main part of the surface energy is due to the isoscalar density-gradient term,
associated with the coefficient $C_{11}^{\nabla}$, which is much smaller for SGII than for SLy230a and RATP:
consequently, large density gradients should cost less energy with SGII.

\begin{table}[t]
\begin{center}
\renewcommand{\arraystretch}{1.5}
\begin{tabular}{|c|c|c|c|c|}
\hline 
& $C_{\rm nn}^{\nabla}$ & $C_{\rm np}^{\nabla}$ & $C_{11}^{\nabla}$ & $C_{33}^{\nabla}$ \\ [-0.15cm]
& \;[MeV.fm$^{5}$] \;& \;[MeV.fm$^{5}$] \;& \;[MeV.fm$^{5}$] \;& \;[MeV.fm$^{5}$] \; \\
\hline 
\;SLy230a\; & $84.5632 $ & $70.8227 $ & $77.693 $ & $6.87028 $ \\
\hline 
SGII & $43.275 $ & $66.3604 $ & $54.8177 $ & $-11.5427$ \\
\hline 
RATP & $80.0409 $ & $79.9703 $ & $80.0056 $ & $0.0353125 $ \\
\hline
\end{tabular}
\end{center}
\caption
{Coefficients of the density-gradient terms in the Skyrme density functional.}
\label{TAB:coef-Hfin}
\end{table}%

The local-Fermi-momentum approximation
generates a coordinate-dependent Fermi momentum $k_{\rm F}(\rv)$.
Together with the temperature $T$, 
it provides expressions for the local particle, kinetic-energy and entropy densities.
Using the Skyrme energy functional and the entropy,
we can directly evaluate the free-energy density,
which can be decomposed into a bulk and a surface term.

The bulk term contains the bulk part of the energy functional 
(Eq.~(\ref{EQ:E_HNM})) and the entropy term.
It is obtained from the space average of
the free energy density $\mathrm{f}^{\rm h}$ (defined in Eq.~(\ref{eq:fh})) 
calculated with the local densities~\cite{CD-A3}.
Performing a development up to the second order in the amplitudes
$\mathrm{A}_i$ associated with the density fluctuations - the first
order being null due to particle conservation - the 
bulk variation of the free-energy is given by:
\begin{equation}
\label{EQ:delta-fb}
\delta \mathrm{f}^{\rm b} =\sum_{i,j}\frac{\mathrm{A}_{i}\mathrm{A}_{j}^{*} + 
\mathrm{A}_i^{*}\mathrm{A}_{j}}{2} \;
\frac{\partial{\mu_i}(\rho_{0,\rm n}, \rho_{0,\rm p})}{\partial \rho_j}
\; ,
\end{equation}
which depends on the curvature-matrix elements of the homogeneous
free-energy $\mathrm{C}^{\rm h}_{ij} = \partial \mu_i/ \partial \rho _j$.

The surface term is generated by the gradient term of the energy functional.
The space average of $\mathcal H^{\nabla}$ leads to the surface contribution:
\begin{equation}
\label{EQ:delta-enabla}
\delta \mathcal E^{\nabla}=\frac{1}{\Omega} \int {\mathcal H^{\nabla}(\rv) \rm d \rv}
= \q^2 \sum_{i,j}\left(\mathrm{A}_i\mathrm{A}_j^*+\mathrm{A}_i^*\mathrm{A}_j \right) 
\mathrm{C}_{ij}^{\nabla}
\; .
\end{equation}

Considering finite-size fluctuations, 
one shall also take into account the effect of the Coulomb interaction.
This was not the case in the thermodynamic framework discussed in Sec.~\ref{SEC:thermo},
since it is well known that the Coulomb energy diverges in infinite charged matter.
In star matter, electrons equilibrate the positive charges and remove the divergence. 
In finite nuclei, the Coulomb energy might be large but finite. 
In both cases, the physically important quantity is not the total Coulomb energy 
but  its variation $\delta\mathcal{E}^{\rm c}$ induced by density fluctuation.
In the present calculation,
we have neglected the small contribution of the exchange Coulomb interaction 
so that the Coulomb contribution to the free-energy variation reads:
\begin{equation}
\label{EQ:delta-ec}
\delta \mathcal E^{\rm c}
=\frac{4\pi e_ie_j}{\q^2} \sum_{i,j} \frac{\mathrm{A}_i\mathrm{A}_j^*+
\mathrm{A}_i^*\mathrm{A}_j }{2} 
\; ,
\end{equation}
where $e_{\rm p}=q_{\rm e}/\sqrt{4\pi\epsilon_0}$, $e_{\rm n}=0$, with $q_{\rm e}$ the elementary electric charge.

Finally, the total free-energy variation is given by the summation of the three terms,
$\delta \mathrm{f} =\delta \mathrm{f}^{\rm b}+\delta\mathcal E^{\nabla}+\delta\mathcal E^{\rm c}$:
\begin{equation}
\label{EQ:delta-f-bis}
\delta \mathrm{f}
=\sum_{i,j} \frac{\mathrm{A}_i\mathrm{A}_{j}^*+\mathrm{A}_i^*\mathrm{A}_{j}}{2}
\left( \frac{\partial{\mu_i}(\rho_{0,\rm n}, \rho_{0,\rm p})}{\partial \rho_j}
+2C_{ij}^{\nabla}\q^2 +\frac{4\pi e_i e_{j}}{\q^2} \right)
\; ,
\end{equation}
which can be expressed in a matricial form in the space of the density fluctuations
$\tilde{\mathrm{A}}=\left( \mathrm{A}_{\rm n}, \mathrm{A}_{\rm p}\right)$:
$\delta \mathrm{f} = {\tilde{\mathrm{A}}^*} \mathrm{C}^{\rm f} \tilde{\mathrm{A}}$, where
\be
\label{EQ:Cf}
\mathrm{C}^{\rm f}= \mathrm{C}^{\rm h}+ 2\q^2 
\left( 
\begin{array}{cc} C_{\rm nn}^{\nabla} & C_{\rm np}^{\nabla}\\ C_{\rm pn}^{\nabla} & C_{\rm pp}^{\nabla}\\ \end{array}
\right) 
+ \frac{4\pi e^2}{\q^2} 
\left(
\begin{array}{cc} 0 & 0\\ 0 & 1\\ \end{array}
\right)
\; .
\ee

The curvature matrix $\mathrm{C}^{\rm f}$ 
is a function of the temperature $T$, the average
densities $\rho_i^0$, and the wave number of the fluctuation $\q$.  
It contains the bulk term of the curvature, $\mathrm{C}^{\rm h}$, 
the density-gradient term of the nuclear interaction proportional to $\q^2$,
and the Coulomb-interaction term proportional to $1/\q^2$.  
These two additional terms being positive tend to reduce the
instability induced by the bulk term.
Thus, the instability will be quenched 
at large values of $\q$ by the surface contribution,
and at small values of $\q$ by the Coulomb interaction.
It is then interesting to evaluate the curvature matrix for intermediate values of $\q$,
where it may still acquire negative values.
In the following, the calculations show that there is still a region where
the lower eigen-value of $\mathrm{C}^{\rm f}$ is negative for a finite interval of $\q$ values,
leading to the development of spinodal instabilities.

\subsection{Hydrodynamical evolution}
\label{SSEC:hydro}

The free-energy curvature analysis presented so far 
provides a static description of the finite-size instabilities.
In order to describe the early dynamics of matter clusterization induced by such instabilities, 
we use a hydrodynamical framework, which corresponds to the local equilibrium limit
and so is consistent with the above Thomas-Fermi approximation of $\delta \mathrm{f}$.
Let us first consider a single fluid described by the 
Euler and continuity equations \cite{PC-PhysRep}:
\begin{eqnarray}
\label{EQ:Euler}
\rm d(m\rho \mathbf{v})/ \rm dt = -\nabla P \; , \\
\label{EQ:continuity}
\partial \rho / \partial t + \nabla (\rho \mathbf{v})=0 \; ,
\end{eqnarray}
where $\mathrm{P}$ is the system's pressure,
$\mathrm{m}$ is the particle mass and $\rho$ the particle density. 
In the limit of small-amplitude fluctuations $\rho=\rho_0+\delta\rho$,
Eqs.~(\ref{EQ:Euler}-\ref{EQ:continuity}) can be transformed into a diffusion equation:
\begin{equation}
\label{EQ:linerarized-euler}
\frac{\partial^2 \rho}{\partial t^2}
=\frac{1}{\mathrm{m}}\; \frac{\rm dP(\rho_0)}{\rm d\rho} \; \nabla^2 \rho
= \frac{\rho_0}{\mathrm{m}} \; \frac{\rm d^2f(\rho_0)}{\rm d\rho^2} \; \nabla^2 \rho \;.
\end{equation}
which is obtained using the relation between the pressure and the
free-energy density $\mathrm{f}$,
$\rm dP/\rm d\rho=\rho \rm d^2f/\rm d\rho^2$.
Expanding Eq.~(\ref{EQ:linerarized-euler}) on a plane-wave basis
gives the following dispersion relation for the density modes:
\begin{equation}
\label{EQ:hydro1F}
\omega^2 = \frac{\rho_0}{\mathrm{m}} \;  \frac{\rm d^2f(\rho_0)}{\rm d\rho^2} \;  \q^2 \;.
\end{equation}
If the free-energy curvature $\rm d^2f/\rm d\rho^2$ is negative, 
the frequency is purely imaginary, $\omega=\rm i\gamma$,
and induces an exponential growth of the fluctuation
with typical time $\tau = 1/\gamma$.  
In asymmetric unstable nuclear matter, we generalize Eq.~(\ref{EQ:hydro1F}) 
introducing fluctuations along the unstable direction of the curvature matrix.
This leads to :
\begin{equation}
\label{EQ:hydro-heuristic}
\omega^2 = - \gamma^2 = \frac{\rho}{\mathrm{m}} \frac{\mathrm{C}^{\rm f}_<}{2} \q^2
\; ,
\end{equation}
where $\mathrm{m}$ is the nucleon mass, $\rho$ the total average density and
$\mathrm{C}^{\rm f}_<$ the lower eigen-value
of the free-energy curvature matrix in the space of density fluctuations.
The factor $1/2$ is due to the change of variables from
$(\rho,\rho_3)$ to $(\rho_{\rm n},\rho_{\rm p})$.
Equation~(\ref{EQ:hydro-heuristic}) characterizes the properties of the finite-size instabilities 
in terms of the free-energy curvature matrix,
in the framework of a semi-classical and local equilibrium approach.
In the following, it will be denoted as the hydrodynamical approach
\footnote
{
It should be noticed that if we are deriving the hydrodynamical equations 
from a microscopic Boltzmann propagation,
while the conservation of the proton and neutron numbers yields two continuity equations,
the conservation of momenta corresponds to a unique Euler equation for the fluid at constant asymmetry $y$.
Indeed, in a proton-neutron collision, only the total momentum is conserved,
and not the proton and neutron ones separately.
Moreover, when the local equilibrium is obtained assuming the divergence of the various cross-sections 
(including the neutron-proton one), the two fluids are forced to follow the same motion 
and thus cannot physically separate.
In this case, the problem reduces to a single fluid dynamics, the asymmetry being fix,
and the instability condition is given by the mechanical condition 
$\partial \mathrm{P}/\partial\rho|_{y=cte}<0$,
Eq.~(\ref{EQ:hydro1F}) becoming $\omega^2=\partial \mathrm{P}/\partial\rho|_{y=cte}\q^2/\mathrm{m}$.
A different assumption is made in the present approach, 
where the instability is characterized by the minimal curvature of the free-energy:
the direction of phase separation in the plane $(\rho_{\rm n},\rho_{\rm p})$ is imposed before treating the time evolution.
}.

\section{Collisionless dynamics of cluster formation}
\label{SEC:collisionless}

Let us now consider a fully quantal approach 
for the description of the growth of density instabilities.
This approach is based on the linear response theory to unstable fluctuations.
In the hydrodynamical approach, a local equilibrium was assumed:
in microscopic words, it means that for each particle species
a Fermi sphere centered on zero momentum was defined in each point of space.
Now, in contrast, each particle species is associated with a single Fermi sphere
affected by a momentum shift (particle-hole excitations) causing density fluctuations in coordinate space.
These two limits correspond respectively to first and zero sound.

The linear response function is obtained from the small-amplitude limit 
of the TDHF theory (Time-Dependent Hartree-Fock):
this corresponds to the RPA (Random-Phase Approximation)~\cite{Ring-Schuck}.
In the present work we use the techniques developed in
Refs.~\cite{gar92,bra95} to obtain the RPA response at finite temperature, 
in the peculiar case of Skyrme-type effective interactions.
We first solve the RPA in a fully quantal framework.
We then introduce the semi-classical limit of the RPA, namely the Vlasov approach,
in order to identify the quantum effects 
in the comparison between hydrodynamics and RPA results.

\subsection{Quantal effects described in the HF+RPA framework}
\label{SSEC:RPA}

In the quantal framework, we have to linearize the TDHF equation, 
which describes the time evolution of the one-body density matrix
$\rho_{ij}=\langle \hat{\mathrm{c}}^{\dag}_j \hat{\mathrm{c}}_i\rangle$, 
$\hat{\mathrm{c}}^{(\dag)}_i$ being the annihilation (creation) operator 
associated with the single-particle state $i$:
\begin{equation}
\label{EQ:TDHF}
{\rm i\hbar} \frac{\partial\hat\rho}{\partial t} = [\hat{\mathrm{h}}+\hat{\mathrm{V}}_{\rm ext}, \hat \rho]
\; ,
\end{equation}
where $\hat{\mathrm{h}}$ is the mean-field operator and $\hat{\mathrm{V}}_{\rm ext}$ is an external field.
In the absence of perturbation, the mean-field ground state is given by 
$\hat\rho_0$ and $\hat{\mathrm{h}}_0$, which is the Hartree-Fock solution
satisfying $[\hat{\mathrm{h}}_0,\hat \rho_0]=0$.
In the following, we consider spin-saturated systems.
For infinite nuclear matter, the operators $\hat{\mathrm{h}}_0$ and $\hat \rho_0$ 
are diagonal in momentum space and we have:
\begin{equation}
\label{ }
\left\{
\begin{array}{l}
\hat \rho_0 |\kv\rangle = g \mathrm{f}_{\kv} |\kv\rangle \\
\hat{\mathrm{h}}_0 |\kv\rangle = \hbar\omega_{\kv} |\kv\rangle
\end{array}
\right.
\end{equation}
where $g$ is the spin degeneracy, 
and $\mathrm{f}_{\kv}$ the Fermi-Dirac occupation for a level of momentum $\kv$.

In this section, we consider a large system of volume $\Omega$ 
and  the plane waves $|\kv\rangle$ are defined
such that $\langle\rv|\kv\rangle={\rm e}^{{\rm i}\kv\cdot\rv}/\sqrt{\Omega}$.

\subsubsection{Linear response to an external probe}
\label{SSSEC:response}

We calculate the response function of nuclear matter at finite temperature
to an infinitesimal external field of the form:
\begin{equation}
\begin{array}{llll}
\hat{\mathrm{V}}_{\rm ext} = \mathcal E {\rm e}^{{\rm i}\qv\cdot \hat{\rv}-{\rm i}(\omega+{\rm i}\eta) t} & (t<0) & ; \; 
\hat{\mathrm{V}}_{\rm ext} = 0 & (t \geq 0)
\; ,
\end{array}
\end{equation}
where $\eta$ is a positive infinitesimal part ensuring the adiabatic
switching of the external field.
The excitation carries a momentum $\qv$ and an energy $\omega$.
For simplicity, we first treat nucleons as a single-component fluid.  
At late enough times and small enough external field $\mathcal E$, the expectation
value of the density fluctuation has the same space and time
dependence as the external field:
\be
\delta\rho(\rv,t)&=&\alpha {\rm e}^{{\rm i}\qv\cdot\rv-{\rm i}(\omega+{\rm i}\eta) t}\;.
\ee
The linearization of Eq.~(\ref{EQ:TDHF}) leads to the polarization function:
\be
\label{EQ:polar}
\Pi(\omega,\qv)= \frac{\alpha}{\mathcal E} = \frac{\Pi_0(\omega,\qv)}{1-\Pi_0(\omega,\qv) \mathrm{v}^{\rm res}} \;.
\ee
In this expression, we have introduced the Lindhard function $\Pi_0$ (corresponding to the response function in the absence of residual interaction): 
\begin{equation}
\label{EQ:Pi0-1}
\Pi_0 (\omega,\qv) 
= \frac{g}{\hbar (2\pi)^3} \int \mathrm{d}\kv \frac {\mathrm{f}_{\kv} - \mathrm{f}_{\kv+\qv}} {\omega -\omega_{\mathbf{kq}} + {\rm i} \eta}
\;\;\; ; \;\;\; \omega_\mathbf{kq}=\omega_{\kv+\qv}-\omega_{\kv}
\;,
\end{equation}
and the residual interaction $\mathrm{v}^{\rm res}$ which accounts for the variation of the mean field
$\delta \hat{\mathrm{h}}$ due to the density fluctuation.
Details for the derivation of the response function~(\ref{EQ:polar}) 
are given in Appendix~\ref{APP:response}.

A collective mode $(\omega,\qv)$ of the nuclear fluid corresponds to a large enhancement of the
response function, or a small value of the denominator of Eq.~(\ref{EQ:polar}).
In the spinodal unstable zone, the response function diverges for a given value of $\omega$ and $\q$.
The dispersion relation $\omega(\qv)$ for the unstable mode is then defined
by $(1-\Pi_0(\omega(\qv),\qv)\mathrm{v}^{\rm res}=0)$.

Let us now take into account the isospin degree of freedom.  
In this case, we distinguish neutron and proton fluids, 
and we get the response function and the residual interaction in terms of the matrices:
\begin{eqnarray}
\mathbf{\Pi_0} &=& \left(\begin{array}{cc}\Pi_{0,{\rm nn}} & 0 \\0 &
\Pi_{0,{\rm pp}}\end{array}\right) \; , \\
\mathbf{v}^{\rm res} &=& \left(\begin{array}{cc}\mathrm{v}^{\rm res}_{\rm nn} & \mathrm{v}^{\rm res}_{\rm np} \\\mathrm{v}^{\rm res}_{\rm pn} &
\mathrm{v}^{\rm res}_{\rm pp}\end{array}\right) \; , \\
\mathbf{\Pi} 
&=&\left(1-\mathbf{\Pi_0}\mathbf{v}^{\rm res}\right)^{-1}\mathbf{\Pi_0} \; .
\label{EQ:RPA}
\end{eqnarray}
The linear response $\mathbf{\Pi}(\omega,\qv)$ presents a divergence 
for the zero of matrix $\left(1-\mathbf{\Pi_0}\mathbf{v}^{\rm res}\right)$,
corresponding to the condition:
\begin{equation}
\label{EQ:RPA-condition1}
\det \left(1-\mathbf{\Pi_0}\mathbf{v}^{\rm res}\right)=0
\; .
\end{equation}
%

\subsubsection{Residual interaction and free-energy curvature}
\label{SSSEC:residual}

We now have to express the residual interaction $\mathbf{v}^{\rm res}$
entering the definition of the response function (\ref{EQ:RPA}).
We give in Appendix~\ref{APP:residual} the steps demonstrating the link between
$\mathbf{v}^{\rm res}$ and the free-energy curvature matrix $\mathcal{C}^{\rm f}$
discussed in Sect.~\ref{SEC:local equilibrium}.
To summarize, considering a density fluctuation
\be
\delta\rho_i(\rv)=\sum_{\qv}\rho_{i,\qv}
&;&
\rho_{i,\qv}=\mathrm{A}_{i,\qv}{\rm e}^{{\rm i}\qv\cdot\rv}
\;\;\;\;\;\; (i={\rm n},{\rm p})
\;,
\ee
we have the average residual-energy variation per unit volume:
\be
\delta\mE^{\rm res}
&=&\sum_{\qv} \sum_{i,j}
\frac{\mathrm{A}_{i;\qv}\mathrm{A}^*_{j;\qv}}{2} \mathrm{v}^{\rm res}_{ij}(\qv) 
\;\;\;\;\;\; (i,j={\rm n},{\rm p})\;.
\ee
The following expression is obtained for the residual-interaction elements
as a function of the free-energy curvature:
\be
\mathrm{v}^{\rm res}_{ij}(\qv)&=&\mathrm{C}^{\rm f}_{ij}(\qv) - 
\frac{\delta_{ij}}{\mathrm{N}_{0,i}} \;,
\ee
where $\mathrm{N}_{0,i}$ is the density of states at Fermi level for the particle species $i$.
This corresponds to the matrix expression:
%
\be
\mathcal{C}^{\rm f} = \left( \mathbf{N_0} \right)^{-1} + \mathbf{v}^{\rm res} 
&;&
\mathbf{N_0} = \left(\begin{array}{cc}\mathrm{N}_{0,{\rm n}} & 0 \\0 & \mathrm{N}_{0,{\rm p}}
\end{array}\right) \; .
\ee

We remind that according to the semi-classical free-energy criterion discussed in the previous section, 
an instability occurs when the matrix $\mathcal{C}^{\rm f}$ has a negative eigen-value.  
The corresponding spinodal region is circled by a frontier for which the lower eigen-value is zero, 
responding to the condition:
\begin{equation}
\label{EQ:SC-condition}
\det \mathcal{C}^{\rm f}=\det \left[ \left( \mathbf{N_0} \right)^{-1} + \mathbf{v}^{\rm res}\right]=0 \;.
\end{equation}
Now, the RPA modes are defined by the condition (\ref{EQ:RPA-condition1}), 
which can be expressed in a closely related form as:
\be
\label{EQ:rpa-condition2}
\det \mathcal C^{\rm RPA}
=\det \left[ \left( \mathbf{N_0} \right)^{-1} + \mathbf{P_0}\mathbf{v}^{\rm res}\right]=0 \; ,
\ee
where
\be
\label{EQ:def-P0}
\mathbf{P_0}(\omega,\qv) = -\left( \mathbf{N_0} \right)^{-1} \mathbf{\Pi_0}(\omega,\qv) \; .
\ee
Note that the matrix $\mathcal{C}^{\rm RPA}$ which has just been introduced, 
in contrast with any curvature matrix, is not symmetric in the general case.

In the RPA framework, the spinodal frontier is defined by the occurrence of a zero transfered energy mode.
For $T=0$, in the limit $\q \rightarrow 0$
the uncorrelated response function then reduces to~\cite{Fetter-Walecka}: 
\begin{equation}
\label{EQ:equiv1}
\lim_{\q \rightarrow 0} \Pi_{0,i} (\omega=0,\qv) = -\mathrm{N}_{0,i} \;.
\end{equation}
According to Eqs.(\ref{EQ:SC-condition}-\ref{EQ:def-P0}),
this implies that, for $T=0$ and $\q \rightarrow 0$,
the condition defining the RPA spinodal frontier reduces to the semi-classical one, 
$\det \mathrm{C}^{\rm f}=0$.
Note however that, taking into account the Coulomb interaction, the instability interval of $\q$ has a lower limit,
so that the RPA and semi-classical spinodal frontiers always remain separated. 

\subsubsection{Unstable modes in the RPA framework}
\label{SSSEC:RPA-unstable}

In the RPA approach, the instability region is characterized by the
occurrence of imaginary modes: $\omega=\omega_{\rm r} + {\rm i}\gamma$.  
The imaginary term ${\rm i}\gamma$ leads to an exponential increase of the fluctuation amplitude 
with the typical time $\tau=1/\gamma$.  
Considering the general expression of $\omega$ in the complex plane,
the uncorrelated response function expressed by 
Eq.~(\ref{EQ:Pi0-1}) can be written (for each particle species $i$):
\begin{eqnarray}
\Pi_{0,i} (\omega,\qv) &=& \frac{g}{\hbar (2\pi)^3} 
\int{{\rm d}\kv \; \frac{ \mathrm{f}_{i,\mathbf{k}} - \mathrm{f}_{i,\mathbf{k}+\mathbf{q}}} {\omega -\omega_{\mathbf{kq}}} } 
=
\frac{2g}{\hbar (2\pi)^3} \int{{\rm d}\kv \; \mathrm{f}_{i,\mathbf{k}}
\frac{\omega_{\mathbf{kq}}}{\omega^2 - \omega_{\mathbf{kq}}^2} }
\;.
\label{EQ:Pi0-2}
\end{eqnarray}
The condition (\ref{EQ:RPA-condition1}) can be satisfied only if the imaginary part of $\mathbf{\Pi_0}$ cancels, 
which according to Eq.~(\ref{EQ:Pi0-2}) implies that $\omega^2$ is real.
Consequently, unstable modes correspond to a purely imaginary energy such that $\omega={\rm i}\gamma$.
The resulting response function reads:
\begin{equation}
\label{EQ:Pi0-3}
 \Pi_{0,i} (\gamma,\qv) = -\frac{2g}{\hbar (2\pi)^3} 
 \int{{\rm d}\kv \mathrm{f}_{i,\mathbf{k}} \frac{\omega_{\mathbf{kq}}}{\gamma^2 + \omega_{\mathbf{kq}}^2} }
 \;.
\end{equation}
At zero temperature, this integral has an analytic solution:
\begin{equation}
\Pi_{0,i} (\gamma,\qv) = -\mathrm{N}_{0,i} \mathrm{P}_{0,i}(\gamma,\qv)
\label{EQ:Pi0-4}
\end{equation}
where, introducing the quantity $w_i^2=\gamma/[\hbar/2\mathrm{m}_i^{*}]$ 
($w_i$ has the dimension of a momentum), $\mathrm{P}_{0,i}$ is given by:
\be
\mathrm{P}_{0,i}(\gamma,\qv)=\mathrm{P}_{0,i}(w_i(\gamma),\qv) = \frac{1}{2k_{{\rm F}_i} \q} \int_0^{ k_{{\rm F}_i}} {k{\rm d}k \ln \left|
\frac{(2k\q+\q^2)^2+w_i^4}{(2k\q-\q^2)^2+w_i^4} \right|} \nonumber
\;.
\ee
Introducing the reduced variables 
$\nu_i=w_i^2/2\q k_{{\rm F}_i}=\mathrm{m}_i^*\gamma/(\q k_{{\rm F}_i})$ and 
$\q_{{\rm F}_i}=\q/2k_{{\rm F}_i}$, we obtain the analytical expression:
\be
\mathrm{P}_{0,i}(\gamma,\qv) &=& \mathrm{P}_{0,i}(\nu_i(\gamma,\q),\q_{{\rm F}_i}(\q)) \nonumber \\
&=& \frac{1}{2} \left\{ 1 + \frac{1+\nu_i^2-\q_{{\rm F}_i}^2}{4\q_{{\rm F}_i}}
\ln \left|\frac{(1+\q_{{\rm F}_i}^2)^2+\nu_i^2}{(1-\q_{{\rm F}_i}^2)^2+\nu_i^2}\right|
-\nu_i \left[ \arctan \left(\frac{1+\q_{{\rm F}_i}}{\nu_i}\right) + 
\arctan \left(\frac{1-\q_{{\rm F}_i}}{\nu_i}\right)\right]\right\}
\label{EQ:Pi0-5}
\end{eqnarray}
which verifies, consistently with Eq.~(\ref{EQ:equiv1}), the condition at the spinodal frontier:
\begin{equation}
\label{EQ:equiv2}
\lim_{\q \rightarrow 0} \mathrm{P}_{0,i}(\gamma=0,\qv) = 1 \;.
\end{equation}

The RPA dispersion relation for the unstable mode $\gamma(\q)$ is deduced from 
Eqs.~(\ref{EQ:rpa-condition2}) and (\ref{EQ:Pi0-5}).
In this quantal approach, for given temperature and densities, 
uniform matter is unstable against a fluctuation of finite momentum $\q$
if  $\det\mathcal C^{\rm RPA}_{(\q)}(\gamma)$ presents a zero 
for a given value of the typical time $\tau=1/\gamma$.
The corresponding vector gives the direction of the growing fluctuation in the density plane.   

At finite temperature, the same procedure can be followed 
but the response function defined by Eq.~(\ref{EQ:Pi0-3}) needs numerical calculation.
Two different methods have been considered: 
one involves the calculation of Fermi integrals~\cite{Fetter-Walecka} 
and the other consists in performing a pondered sum of zero temperature response functions~\cite{gar92}: 
\be
\rmRe \Pi_{0,i}(\gamma,\q,T)
=-\hspace{-0.12cm}\int_{k}\hspace{-0.12cm} \rmRe\Pi_{0,i}(\gamma,\q,T=0,k_F=k) {\rm d}\mathrm{f}_i(k,T) \; . 
\ee
At zero temperature, ${\rm d}\mathrm{f}_i(k)=-\delta(k-k_{{\rm F}i}){\rm d}k$, yielding the well-known expression for $\rmRe\Pi_{0,i}$.
Both methods have been used in order to check the accuracy of our results at finite temperature.

\subsection{Vlasov approach}
\label{SSEC:Vlasov}

We now introduce the semi-classical limit of the RPA.
The Vlasov approach describes the matter response to a momentum transfer $\qv$ and energy transfer $\omega$ like the RPA approach,
but in a semi-classical approximation.
It is then an intermediate description between the hydrodynamics and the full quantal RPA.
Again, for simplicity, we give expressions for the case of a single fluid.
The semi-classical version of the TDHF equation is obtained in the limit 
$\hbar \rightarrow 0$~\cite{Ring-Schuck}.
Using a Wigner transform, the commutator divided by $i\hbar$ is then replaced by a Poisson bracket,
where the one-body density and potential operators are expressed
by functions of time, position and momentum $(t,\rv,\kv)$.
This gives the Vlasov equation, describing the semi-classical time evolution in the collisionless limit:
\be
\frac{\partial \mathrm{f}(t,\rv,\kv)}{\partial t}=
\left\{ \mathrm{h}(t,\rv,\kv)+\mathrm{V}_{\rm ext}(t,\rv,\kv)\; , \; \mathrm{f}(t,\rv,\kv) \right\}
\;,
\ee
where $\mathrm{f}$ is the occupation function in phase space, 
$h$ the mean field and $\mathrm{V}_{\rm ext}$ an external potential.
In the unperturbed state, the mean field is $\mathrm{h}_0(\kv)$
and the occupation at temperature $T=1/\beta$ is given by the Fermi-Dirac distribution 
$\mathrm{f}_0(\kv)=\left[1+\exp(\beta(\mathrm{h}_0(\kv)-\mu))\right]$.
The density fluctuation corresponds to the occupation variation 
$\mathrm{f}(t,\rv,\kv)=\mathrm{f}_0(\kv)+\delta \mathrm{f}(t,\rv,\kv)$, 
inducing the mean-field variation 
$\mathrm{h}(t,\rv,\kv)=\mathrm{h}_0(\kv)+\delta \mathrm{h}(t,\rv,\kv)$.

Let us stress here the difference between Thomas-Fermi and Vlasov approaches.
Both approaches give a semi-classical description of the state
occupations, using the function $\mathrm{f}(t,\rv,\kv)$: 
however, they define the perturbation $\delta \mathrm{f}(t,\rv,\kv)$ in different ways.
In the Thomas-Fermi approach, the variation of the local density is associated with a variation of the chemical potential,
according to the local equilibrium description, 
and the occupation function is given by a Fermi-Dirac distribution:
$\mathrm{f}(t,\rv,\kv)=\left[1+\exp(\beta(\mathrm{h}_0(\kv)-\mu(t,\rv)))\right]$.
In terms of the density matrix, this corresponds to a variation of the Fermi surface.
In contrast, in the Vlasov formalism $\mathrm{f}(t,\rv,\kv)$ is given
by a canonical transformation of $\mathrm{f}_0(\kv)$: 
the variation of the distribution can be expressed in terms of a
generating function $\mathrm{S}(t,\rv,\kv)$  
such that $\delta \mathrm{f}=\{\mathrm{S},\mathrm{f}_0\}$~\cite{Providencia-PRC73}. 
In terms of the density matrix, this corresponds to the excitation of particle-hole states
with a finite momentum transfer.

The linearized Vlasov equation (semi-classical limit of the RPA) 
can be solved following a method similar to the one exposed in Ref.~\cite{Providencia-PRC73}.
The Vlasov eigen-modes for asymmetric nuclear matter obey the matrix relation:
\be
\label{EQ:Vlasov-condition}
\det \mathcal C^{\rm Vlasov}(\omega,\qv)
=\det \left[ \left( \mathbf{N_0} \right)^{-1} +\frac{\mathbf{L}\mathbf{v}^{\rm res}}{2}\right]=0 \; ,
\ee
where (with $i={\rm n},{\rm p}$):
\be
\mathbf{L}(s) = \left(\begin{array}{cc}\mathrm{L}(s_{\rm n}) & 0 \\0 & \mathrm{L}(s_{\rm p})\end{array}\right)\; ;
\;\;\;
s_i = \omega\left[\q\frac{{\rm d}\mathrm{h}_{0i}(\{k_i=k_{{\rm F}i}\})}{{\rm d}k}\right]^{-1} \; ;
\;\;\;
\mathrm{L}(s_i) = 2-s_i\ln\left|\frac{1+s_i}{1-s_i}\right|
\; ;
\ee
$\mathrm{L}(s_i)$ denotes the semi-classical Lindhard function.
The low-$\q$ limit of the uncorrelated response function $\Pi_{0,i}(\omega,\qv)$
is related to $\mathrm{L}(s_i)$ in such way that we get in terms of matrixes:
\be
\lim_{\q \rightarrow 0} (-\mathbf{N_0}\mathbf{\Pi_0}(\omega,\qv))
= \lim_{\q \rightarrow 0} \mathbf{P_0}(\omega,\qv)
= \mathbf{L}/2
\; .
\ee
This means that the RPA eigen-modes obeying Eq.~(\ref{EQ:rpa-condition2})
reduce to the Vlasov ones in the low-$\q$ limit (whatever the value of $\omega$).
On the other hand, for $\omega=0$ we have $L(0)/2=1$,
so that $\mathcal C^{\rm Vlasov}(\omega=0,\qv)$ reduces to the
free-energy curvature matrix $\mathrm{C}^{\rm f}(\qv)$: 
consequently, the frontier of the spinodal region is the same for the two semi-classical approaches.
On this common frontier, the Vlasov and hydrodynamics eigen-modes are exactly identical.
Differences shall appear inside the spinodal region where $\omega\neq 0$.

\section{Discussion of the results}
\label{SEC:discussion}

In this section, we present the properties of matter clusterization 
obtained from the different approaches presented above.
The quantal RPA is compared to two semi-classical approaches:
Vlasov and hydrodynamics,
corresponding respectively to the collisionless (zero sound) and local equilibrium (first sound) limits.
We first consider instability against density fluctuations for
several momenta $\q$, exploring the dependence on the temperature and on the nuclear interaction. 
Then, we concentrate on the most unstable mode defined as the mode for
which the typical instability time is minimum and explore the clusterization properties.

\subsection{Unstable eigen-modes}
\label{SSEC:res-modes}

\begin{figure}[t]
\begin{center}
\includegraphics[width=0.8\linewidth]{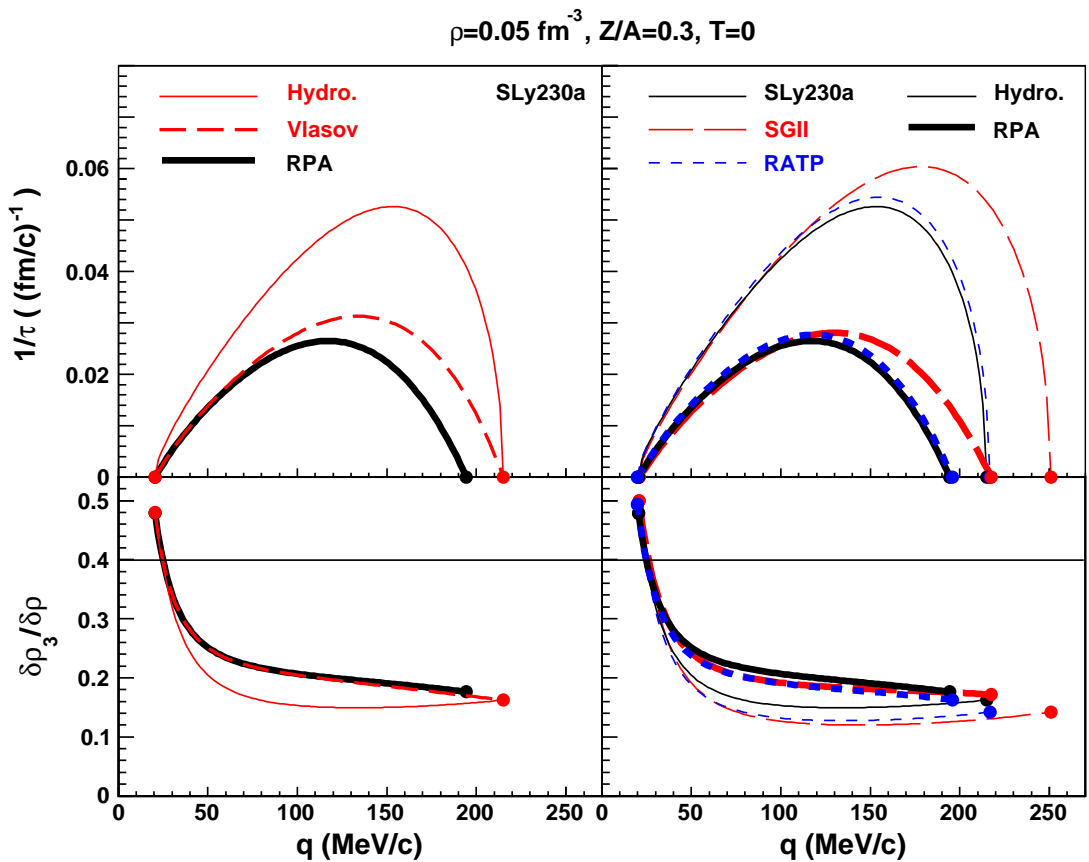}
\caption{(color online) 
	Dispersion relation at a given density and asymmetry, for zero temperature.
	Dots represent the boundaries of the unstable interval of $\q$.
	Top: $1/\tau$ is given as a function of the wave number $\q$.
	Bottom: associated directions of phase separation in the
	density plane $\delta \rho_3/\delta\rho$.  The horizontal line
	at $0.4$ is the direction of constant proton fraction $\rho_3/\rho$.
	Left: three approaches applied with SLy230a:
	RPA (thick line), Vlasov (dashed line), hydrodynamics (thin line).
	Right: RPA and hydrodynamics with different Skyrme forces.}
\label{FIG:disp}
\end{center}
\end{figure}

Let us determine the different instability properties associated with a given wave-number $\q$.
The unstable eigen-modes obey the dispersion relations
illustrated for fixed densities on Fig.~\ref{FIG:disp}.
The upper part of this figure represents the inverse typical time $1/\tau(\q)$,
which characterizes the time evolution of the modes.
The quantum effects can be directly appreciated from the comparison between RPA and Vlasov results.
Although these results are quasi-similar at the low-$\q$ limit,
as expected, they separate for higher values of $\q$.
We then observe a quantum quenching of the instability,
as discussed in Ref.~\cite{Ayik-PLB353}:
the RPA unstable modes are associated with lower values of $1/\tau$
(\emph{i.e.} a slower amplification of the fluctuation)
and they end at a lower value of $\q$.
Indeed, in the Vlasov approach the uncorrelated response function
is approximated by its limit at low $\q$, 
which amounts to neglecting a part of the kinetic-energy cost of the fluctuation.
In the RPA where the exact value of $\Pi_0(\omega,\qv)$ is calculated,
the high-$\q$ density fluctuations are disfavored.
Let us now consider the semi-classical approach based on local equilibrium (hydrodynamics).
It presents the same qualitative features as the previous ones:
in particular, as discussed earlier, the instability region 
corresponding to a negative value of the free-energy curvature
is formally the same as for the Vlasov approach.
However, the hydrodynamical dispersion relation calculated from the
curvature $\mathrm{C}^{\rm f}_<$ 
leads to a much faster evolution, $1/\tau$ being typically twice as large as for the RPA.
It also presents a less symmetric behavior, with a sharper slope toward the high-$\q$ border:
this is due to the relation $1/\tau^2\propto |\mathrm{C}^{\rm f}_<|\q^2$,
in which the factor $\q^2$ enhances the high-$\q$ side of the dispersion relation.

The lower part of Fig.~\ref{FIG:disp} deals with the direction of phase separation, 
also included in the dispersion relation determining the eigen-modes.
This feature is connected with the isospin distillation, 
which is a well-known aspect of the liquid-gas phase transition 
in asymmetric nuclear matter~\cite{Muller-Serot-PRC52}: 
in order to minimize the symmetry energy, the dense phase is more symmetric than the dilute one.
This is observed in the construction of the thermodynamic phase equilibrium, 
and reflected by the spinodal instability direction:
instead of following a line of constant composition in the density plane, 
phase separation is rotated towards the isoscalar direction.
For a neutron rich point of densities $(\rho,\rho_3)$, 
for which phase separation direction is given by the vector
coordinates $(\delta\rho,\delta \rho_3)$,  
the normal isospin distillation is reflected in the relation: 
\begin{equation}
\label{EQ:normal-dist}
0<\delta \rho_3/\delta \rho < \rho_3/\rho
\end{equation}
We have checked that this relation is verified by the thermodynamic
spinodal instability for all the Skyrme-like interactions we consider in this article. 
Moreover, it has been shown in Refs.~\cite{bar06} that for a set of realistic Skyrme-like and Gogny interactions, 
the instability is still isoscalar-like, 
which means that $\delta\rho_3/\delta\rho <0.5$ in a domain of reasonable isospin asymmetries.
Now when finite-size instabilities are considered and the electric charge comes into play, 
the Coulomb interaction disfavors proton-density fluctuations:
this goes against the normal distillation effect (for neutron-rich matter).  
As can be seen on Fig.~\ref{FIG:disp}, isospin distillation is particularly affected at low values of $\q$, 
where Coulomb effects dominate: an inverse distillation is even found 
near the low-$\q$ border of the instability interval.
However, the normal isospin distillation remains the dominant feature, 
especially if we consider the most unstable modes~\cite{Providencia-PRC73,CD-A3}: 
it leads to the formation of clusters more symmetric than the homogeneous initial system, 
immersed in a more neutron-rich gas.

Figure~\ref{FIG:disp} shows a very close agreement between Vlasov and RPA results
as for the value of $\delta\rho_3/\delta\rho$.
The direction corresponding to the hydrodynamical approach
is nothing but the direction of minimal free-energy curvature,
\emph{i.e.} the eigen-vector $\uv_<$ associated with the lower eigen-value 
of the curvature matrix $\mathrm{C}^{\rm f}$. 
Qualitatively, it presents a similar evolution;
furthermore, once again hydrodynamics and Vlasov curves 
are constrained to join at the borders of their common instability interval.
However, inside this interval the hydrodynamics curve is clearly distinct
from the two collisionless approaches.
This shows that the phase separation direction followed in a collisionless approach
is quantitatively different from the direction $\uv_<$
of the negative free-energy curvature:
it leads to a slightly reduced distillation effect.

The right column of Fig.~\ref{FIG:disp} shows 
the dependence of the dispersion relation on the Skyrme force. 
RPA and hydrodynamics results are given for SLy230a, SGII and RATP.
All present similar features, 
ensuring that the results elsewhere shown with SLy230a alone are qualitatively representative.
The quantitative differences between SLy230a and RATP are weak.
SGII however favors modes of larger $\q$-values, 
and presents an extended instability interval on the high-$\q$ side.
This difference comes from the surface properties of the forces:
as can be seen in Tab.~\ref{TAB:coef-Hfin},
the energy cost for the introduction of a density gradient is lower with SGII.

\begin{figure}[t]
\begin{center}
\includegraphics[width=0.8\linewidth]{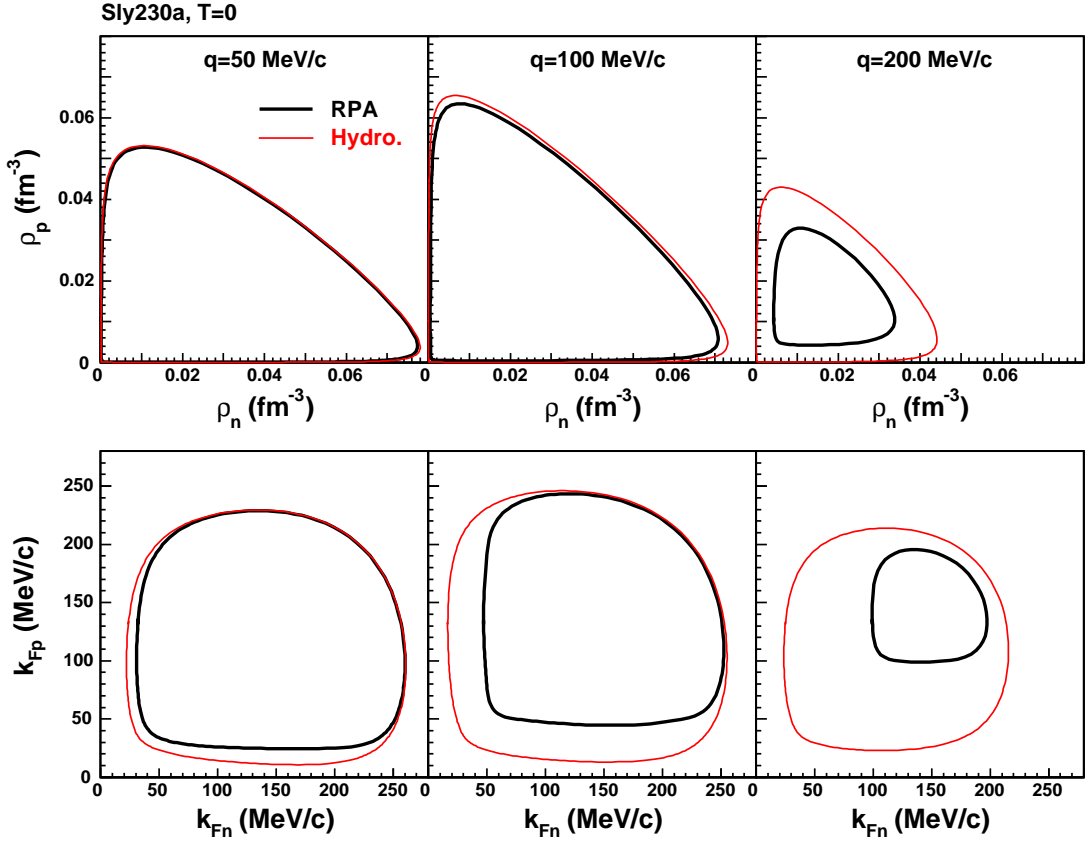}
\caption{(color online) 
	Instability regions associated with fixed values of the 
	wave-number $\q$, at zero temperature, 
	according to RPA (thick curves) 
	and semi-classical (thin curves) criterions.
	Upper part: representation in the density plane.  
	Lower part: representation in the plane of Fermi momenta.
	}
\label{FIG:kspino}
\end{center}
\end{figure}

\begin{figure}[t]
\begin{center}
\includegraphics[width=0.8\linewidth]{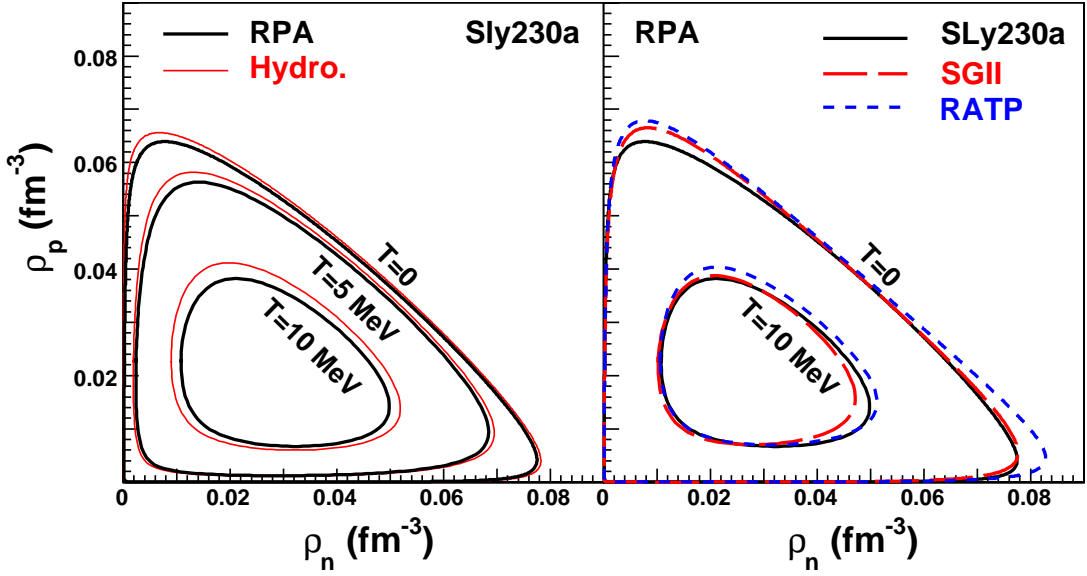}
\caption{(color online) 
	Envelopes of instability against finite-size density fluctuations.
	Left: comparison between semi-classical and RPA criterions, 
	for temperatures $T=0$; $5$; $10$~MeV. 
        	Right: comparison between the RPA instability envelopes
	obtained with different Skyrme forces, for $T=0$; $10$~MeV.
	}
\label{FIG:enveloppes}
\end{center}
\end{figure}

\begin{figure}[t]
\begin{center}
\includegraphics[width=1\linewidth]{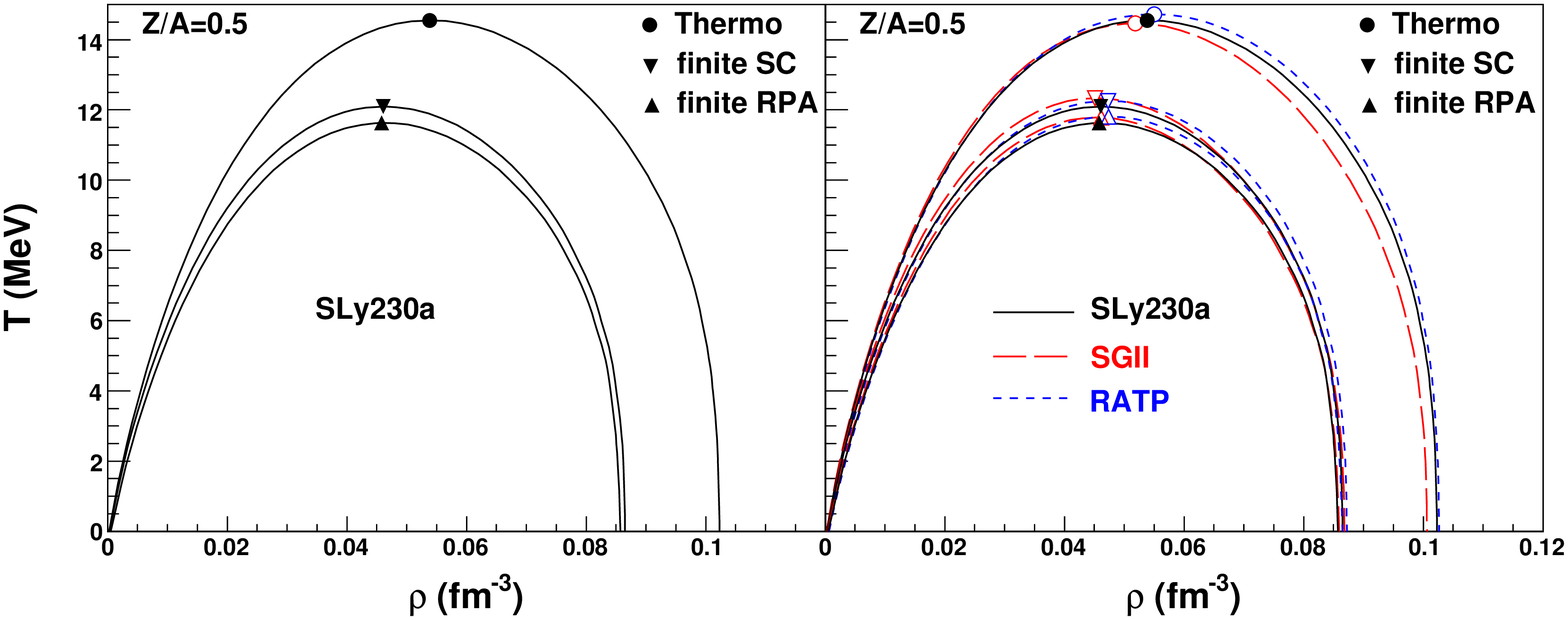}
\caption{(color online) 
	Total-density limits of the instability region depending on the temperature,
	along the axis of symmetric nuclear matter ($Z/A=0.5$).
	The thermodynamic instability region is represented,
	as well the finite-size instability regions obtained in the semi-classical (SC) and RPA approaches.
	The symbols indicate the limiting temperature $T^{\rm s}_{\rm lim}$ in symmetric matter: 
	for the thermodynamic curve, it corresponds to the critical point.
	Left: the three curves obtained with SLy230a.
	Right: comparison of the different Skyrme parameterizations.
	}
\label{FIG:rT-SNM}
\end{center}
\end{figure}

Let us now address the extension of the instability in the density plane.
We first consider instability regions associated with given values of
the transfered-momentum $\q$. 
In a condensed way, we name such regions "$\q$-spinodals".
Inside a $\q$-spinodal contour, a density fluctuation of momentum $\q$
is unstable and amplified according to the dispersion relation discussed above.
Within the RPA approach, the $\q$-spinodal contour at temperature $T$
is determined by the divergence of the linear response $\Pi(\omega=0,\q,T)$.
In the semi-classical description (for both Vlasov and hydrodynamics), 
the $\q$-spinodal contour corresponds to a zero eigen-value of
$\mathrm{C}^{\rm f}(\q,T)$. 
As shown by Eq.~(\ref{EQ:equiv1}), both curves are equivalent at the limit $\q \rightarrow 0$.
Figure~\ref{FIG:kspino} represents different RPA and semi-classical $\q$-spinodals at $T=0$.
For $\q=50$~MeV/c almost no difference can be seen in the density plane.  
On the lower part of the figure, the $\q$-spinodals are drawn in the plane of Fermi momenta, 
$k_{{\rm F}_i}=(3\pi^2\rho_i)^{1/3}$: we can then check that both approaches become very close for $\q \ll k_{{\rm F}_i}$.  
For higher values of $\q$, the RPA criterion always leads to a
reduction of instability compared with the semi-classical approach,
consistently with the previous discussion of the 
quantum quenching affecting the dispersion relation (Fig.~\ref{FIG:disp}).

Figure~\ref{FIG:enveloppes} represents the global region of instability against clusterization.
For each approach, and for a given temperature, 
it corresponds to the envelope of all the $\q$-spinodals:
inside this envelope, nuclear matter can be found unstable 
against finite-size fluctuations for at least one $\q$-value. 
The right par of Fig.~\ref{FIG:enveloppes} compares the RPA instability envelopes 
for different Skyrme-type forces: the slight differences obtained 
reflect those appearing between the thermodynamic spinodal curves of Fig.~\ref{FIG:spino_rnrp_3F}.
Since the widest $\q$-spinodal contour is obtained for $\q$ around $100$~MeV/c, 
the corresponding graph of Fig.~\ref{FIG:kspino} is representative of the difference 
between the global instability regions obtained in RPA and semi-classical approaches at $T=0$.
On the left part of Fig.~\ref{FIG:enveloppes} are shown 
the instability regions obtained at $T=5$ and $10$~MeV for both approaches.
The difference between the two envelopes is sensitively enhanced 
at higher temperature, as we approach the limiting temperature of the instability:
this means that this limit occurs earlier in the RPA due to the quantum quenching.
It is then clear that the temperature range concerned by the liquid-gas instabilities
is not sufficient for a suppression of the quantum effects.

The temperature dependence of the instability regions is illustrated 
in Fig.~\ref{FIG:rT-SNM} for symmetric matter,
considering the thermodynamic and finite-size (semi-classical and RPA) spinodals.
We can observe the reduction of the thermodynamic instability,
and the smaller temperature range of the RPA instabilities with respect to the semi-classical ones.
The temperature and density coordinates of the highest-temperature points 
appearing on this figure are reported in Tab.~\ref{TAB:TlimSym} for the different Skyrme forces.
In the thermodynamic case, the phase-transition region ends in a critical point occurring for symmetric matter; 
its coordinates are $(\rho_{\rm c},T_{\rm c})$.
In contrast, the finite-size instability region does not reach its limiting temperature 
on the axis of symmetric matter, 
the charge-exchange symmetry being broken by the Coulomb interaction.
Instead, the limiting point is given by the coordinates 
$(\rho_{\rm lim},(Z/A)_{\rm lim},T_{\rm lim})$
(associated with the last unstable momentum $\q_{\rm lim}$).
These coordinates are given in Tab.~\ref{TAB:Tlim}, in the semi-classical and RPA cases,
for the different Skyrme forces.
Let us first compare the RPA and semi-classical limiting points.
The limiting temperature is reduced by several hundreds of keV
due to the quantum quenching of the instability in the RPA approach.
The position of the limiting point is found to be at slightly higher density for the RPA:
however, this feature is not very pronounced and is not systematic at given $Z/A$,
as can be seen on Fig.~\ref{FIG:rT-SNM} for $Z/A=0.5$. 
However, it is noticeable that the RPA limiting point occurs at higher asymmetry:
this can be attributed to the larger influence of Coulomb effects,
since, as expected, the corresponding transfered momentum is lower.
Let us now compare the critical and limiting temperatures.
For the forces with similar surface properties 
(\emph{e.g.}, in the presented results, SLy230a and RATP), 
the difference between the values of $T_{\rm lim}$
reflects the difference between the thermodynamical critical temperatures $T_{\rm c}$
(typically reduced by $2.4$ and $2.8$~MeV respectively 
in the semi-classical and RPA cases).
A smaller reduction of the critical temperature is observed for the force with reduced surface energy
(SGII), for which the last unstable momentum $\q_{\rm lim}$ is higher.

\begin{table}[t]
\begin{center}
\renewcommand{\arraystretch}{1.5}
\begin{tabular}{|c|c|c|c|c|c|c|}
\hline 
& $\rho_{\rm c}$ & $\rho_{\rm lim}^{\rm s,SC}$ & $\rho_{\rm lim}^{\rm s,RPA}$ & $T_{\rm c}$ & $T_{\rm lim}^{\rm s,SC}$ & $T_{\rm lim}^{\rm s,RPA}$ \\
[-0.15cm]
& \;[fm$^{-3}$]\; & \;[fm$^{-3}$]\; & \;[fm$^{-3}$]\; & \;[MeV]\; & \;[MeV]\; & \;[MeV]\; \\
\hline
\;Sly230a\;
&$0.0538$&$0.0461 $&$0.0458 $&$14.55$&$12.09 $&$11.63 $\\
\hline
SGII
&$0.0518$&$0.0451 $&$0.0461 $&$14.46$&$12.33 $&$11.78 $\\
\hline
RATP
&$0.0550$&$0.0473 $&$0.0474 $&$14.72$&$12.25 $&$11.81 $\\
\hline
\end{tabular}
\end{center}
\caption
{
Limiting density and temperature for symmetric nuclear matter (subscript "s").
The thermodynamic critical point is compared with the values obtained from the semi-classical (SC) and RPA approaches.
}
\label{TAB:TlimSym}
\end{table}%

\begin{table}[t]
\begin{center}
\renewcommand{\arraystretch}{1.5}
\begin{tabular}{|c|c|c|c|c|c|c|c|c|c|c|}
\hline 
& $\rho_{\rm c}$ & $\rho_{\rm lim}^{\rm SC}$ & $\rho_{\rm lim}^{\rm RPA}$ & $\;(Z/A)_{\rm lim}^{\rm SC}\;$ & $\;(Z/A)_{\rm lim}^{\rm RPA}\;$ 
& $T_{\rm c}$ & $T_{\rm lim}^{\rm SC}$ & $T_{\rm lim}^{\rm RPA}$ & $q_{\rm lim}^{\rm SC}$ & $q_{\rm lim}^{\rm RPA}$ \\
[-0.15cm]
& \;[fm$^{-3}$]\; & \;[fm$^{-3}$]\; & \;[fm$^{-3}$]\; &  &  
& \;[MeV]\; & \;[MeV]\; & \;[MeV]\; & \;[MeV/c]\; & \;[MeV/c]\; \\
\hline
\;Sly230a\;
&$0.0538$&$0.0467 $&$0.0474$&$0.426 $&$0.416$
&$14.55$&$12.20 $&$11.77$&$76.7 $&$70.1$\\
\hline
SGII
&$0.0518$&$0.0456 $&$0.0460$&$0.447 $&$0.431$
&$14.46$&$12.39 $&$11.87$&$85.0 $&$75.1$\\
\hline
RATP
&$0.0550$&$0.0478 $&$0.0478$&$0.436 $&$0.427$
&$14.72$&$12.34$&$11.92$&$76.6 $&$70.2$\\
\hline
\end{tabular}
\end{center}
\caption
{Thermodynamic critical point (total density $\rho_{\rm c}$, 
proton fraction $Z/A=0.5$ and temperature $T_{\rm c}$)
compared with the limiting values obtained from
semi-classical (SC) and RPA for finite-size instabilities.
}
\label{TAB:Tlim}
\end{table}%

\begin{figure}[t]
\begin{center}
\includegraphics[width=0.9\linewidth]{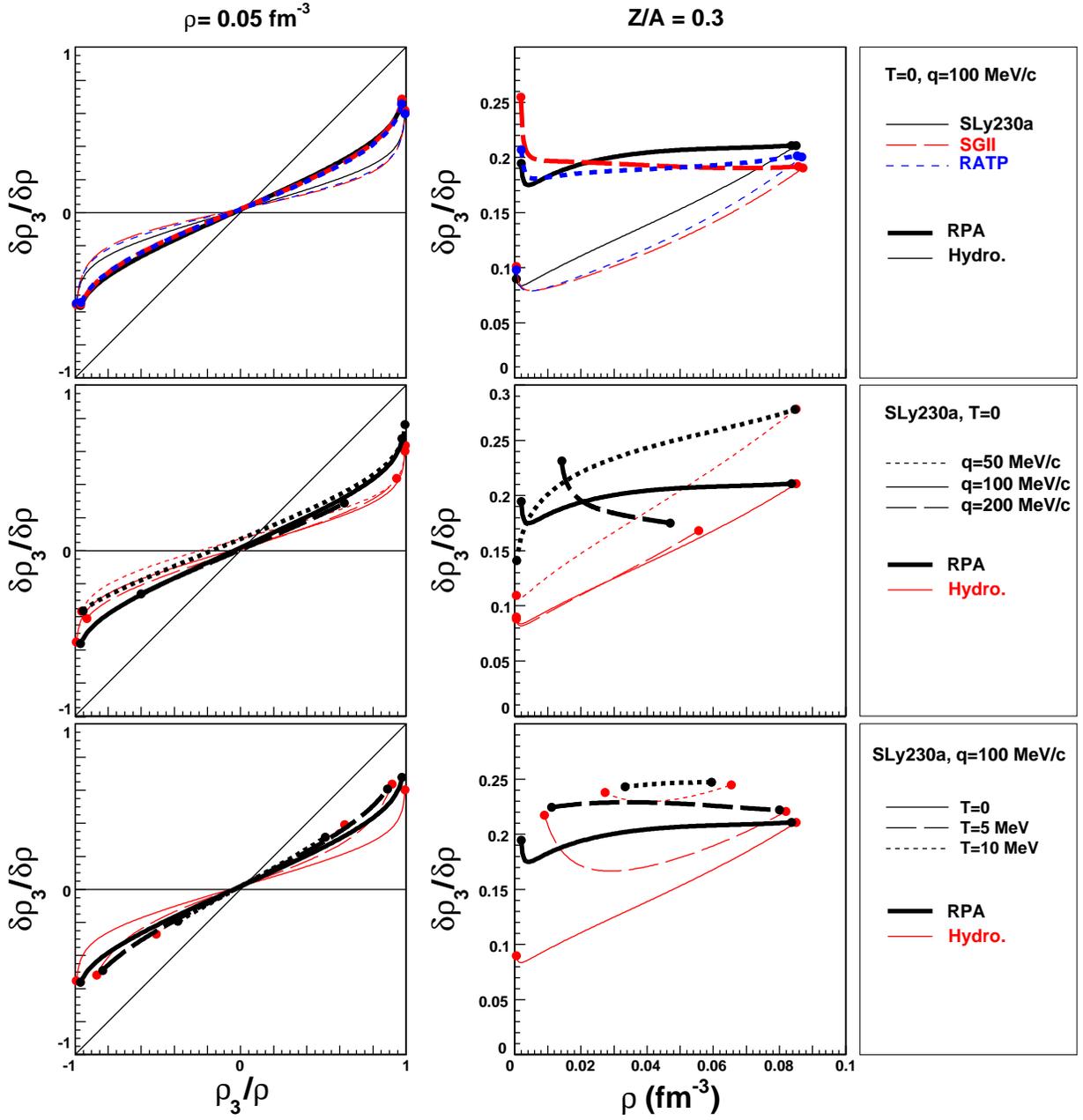}
\caption{(color online) 
	Directions of phase separation for fixed values of the wave number $\q$ 
	studied along two different axis of the density plane, 
	for the hydrodynamical (thin lines) 
	and RPA (thick lines) approaches.
	Dots give the limit of instability regions.
	Left: constant density.
	The horizontal line corresponds to the isoscalar direction, 
	and the first diagonal to the constant-$Z/A$ direction.
	Right: constant proton fraction.
	Top figures compare different Skyrme forces; 
	middle figures different values of $\q$; 
	bottom figures different temperatures.  }
\label{FIG:qcte}
\end{center}
\end{figure}

We finally explore on Fig.~\ref{FIG:qcte} the behavior of the phase separation direction $\delta\rho_3/\delta\rho$
associated with the unstable eigen-modes, for discrete values of the momentum $\q$.
RPA and hydrodynamics results are compared.
The view-graphs represented on this figure are separated into two groups: 
on the left-hand side, the total density is fixed at $\rho=0.05$~fm$^{-3}$
and the horizontal axis explores the isospin asymmetry $\rho_3/\rho$;
on the right-hand side, we fix the proton fraction $Z/A=0.3$ (\emph{i.e.} $\rho_3/\rho=0.4$),
and the horizontal axis explores the total density.
In both cases, the instability direction is represented for different Skyrme forces (top),
different $\q$ values (middle),
and different temperatures (bottom).
First, it could be remarked from the left view-graphs 
that the different curves are mostly situated inside the cone 
defined by the horizontal line and the first diagonal, 
satisfying the inequalities~(\ref{EQ:normal-dist}).
This confirms the predominance of the normal isospin distillation, already noticed about Fig.~\ref{FIG:disp}.

Let us now consider the right panels of Fig.~\ref{FIG:qcte}.
As expected, the difference between RPA and hydrodynamics results tends to disappear 
at the high-density border of the unstable zone,
as the Fermi momenta become large enough with respect to the transfered momentum $\q$.
Indeed, in this case the RPA reduces to the Vlasov approach, 
which identifies with the hydrodynamics on the borders of the instability interval.
Inside the spinodal zone, the differences are enhanced, especially at $T=0$:
in the hydrodynamical approach, the instability direction $\delta \rho_3/\delta\rho$ strongly decreases 
towards lower densities, while in the RPA approach its density dependence is weak.
As a consequence, the RPA values of $\delta \rho_3/\delta\rho$ are larger (up to a factor 2).
This accentuates the observation made on Fig.~\ref{FIG:disp}, that the distillation effect is reduced in the RPA framework.
However, both approaches again confirm the isoscalar behavior of the unstable mode.
We finally notice that the differences are smoothed for the results at finite temperature:
this is essentially due to the reduction of the instability, 
limiting the range of directions leading to a negative free-energy curvature,
rather than to the disappearance of the quantum effects.
Indeed, we know from the comparison between RPA and Vlasov curves on Fig.~\ref{FIG:disp} 
that these effects do not affect the separation direction at least for $\rho\geq0.05$~fm$^{-3}$.
Furthermore, as discussed previously, quantum effects are still present
in all the range of temperature concerned by the liquid-gas instabilities.

\subsection{Dynamical properties of cluster formation}
\label{SSEC:res-clusters}

In the spinodal decomposition scenario,
the early dynamics of cluster formation is induced by the most unstable mode,
corresponding to the fastest amplification of the fluctuation.
For fixed temperature and densities, this is given by the maximum of the dispersion relation $1/\tau(\q)$,
that we denote  ($\q_0,1/\tau_0$).
Clusters should then be formed according to the timescale $\tau_0$,
with a size of the order of $\lambda_0/2=\pi/\q_0$,
and a composition related to the corresponding direction $(\delta\rho_3/\delta\rho)_0$.

\begin{figure}[t]
\begin{center}
\includegraphics[width=0.8\linewidth]{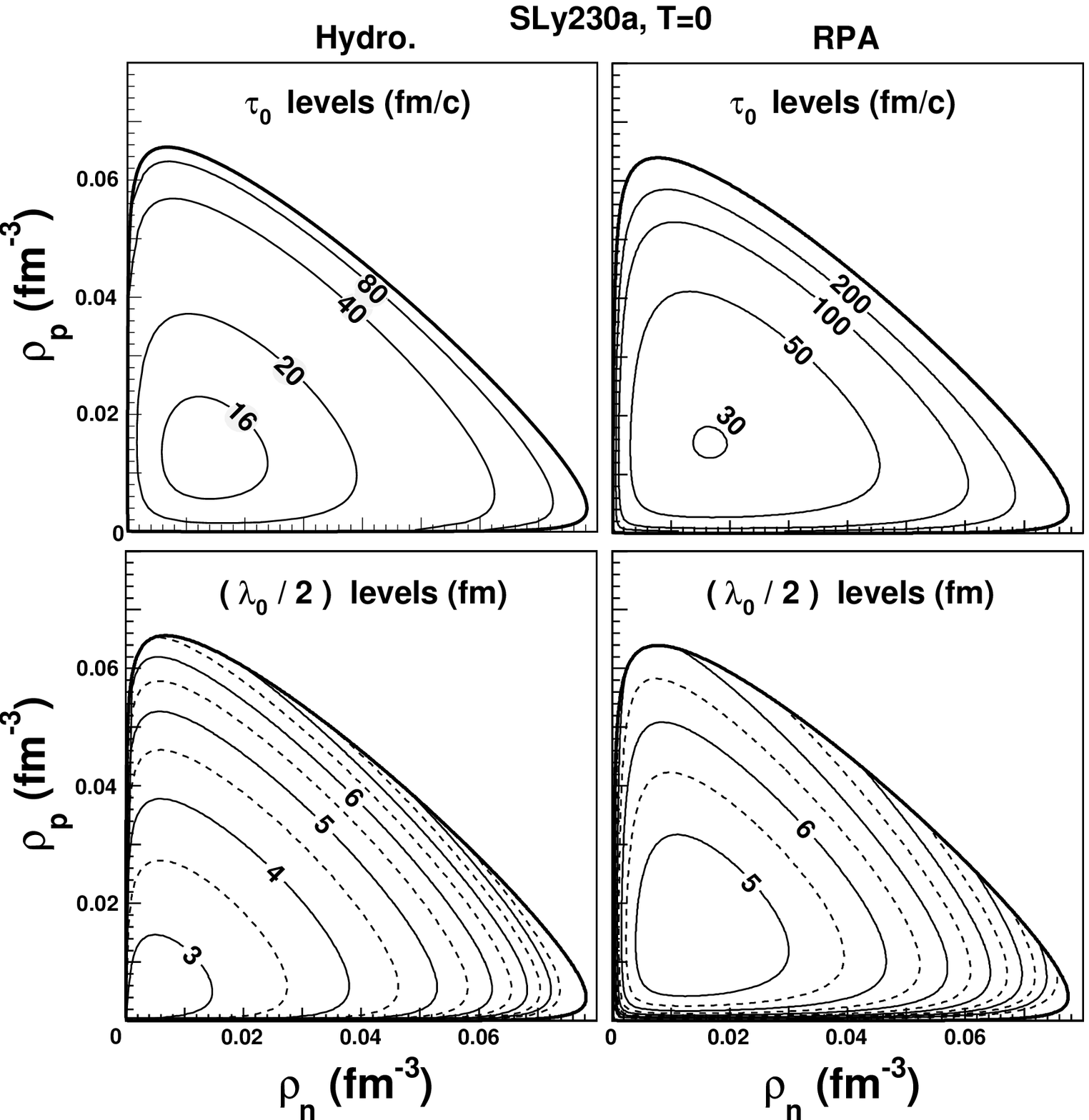}
\caption{(color online) 
	Time and space characteristics of the most unstable mode,
	represented in level-lines inside the instability envelope (thick curve).  
	RPA and hydrodynamics results are shown.  
	Top: time constant $\tau_0$ giving the fastest amplification of the density fluctuation.
	Bottom: half-wavelength $\lambda_0/2=\pi/\q_0$ of the dominant fluctuation mode.  
	Levels are separated by an interval of $0.5$~fm.
	}
\label{FIG:rnrp_tau0l0}
\end{center}
\end{figure}

\begin{figure}[t]
\begin{center}
\includegraphics[width=0.8\linewidth]{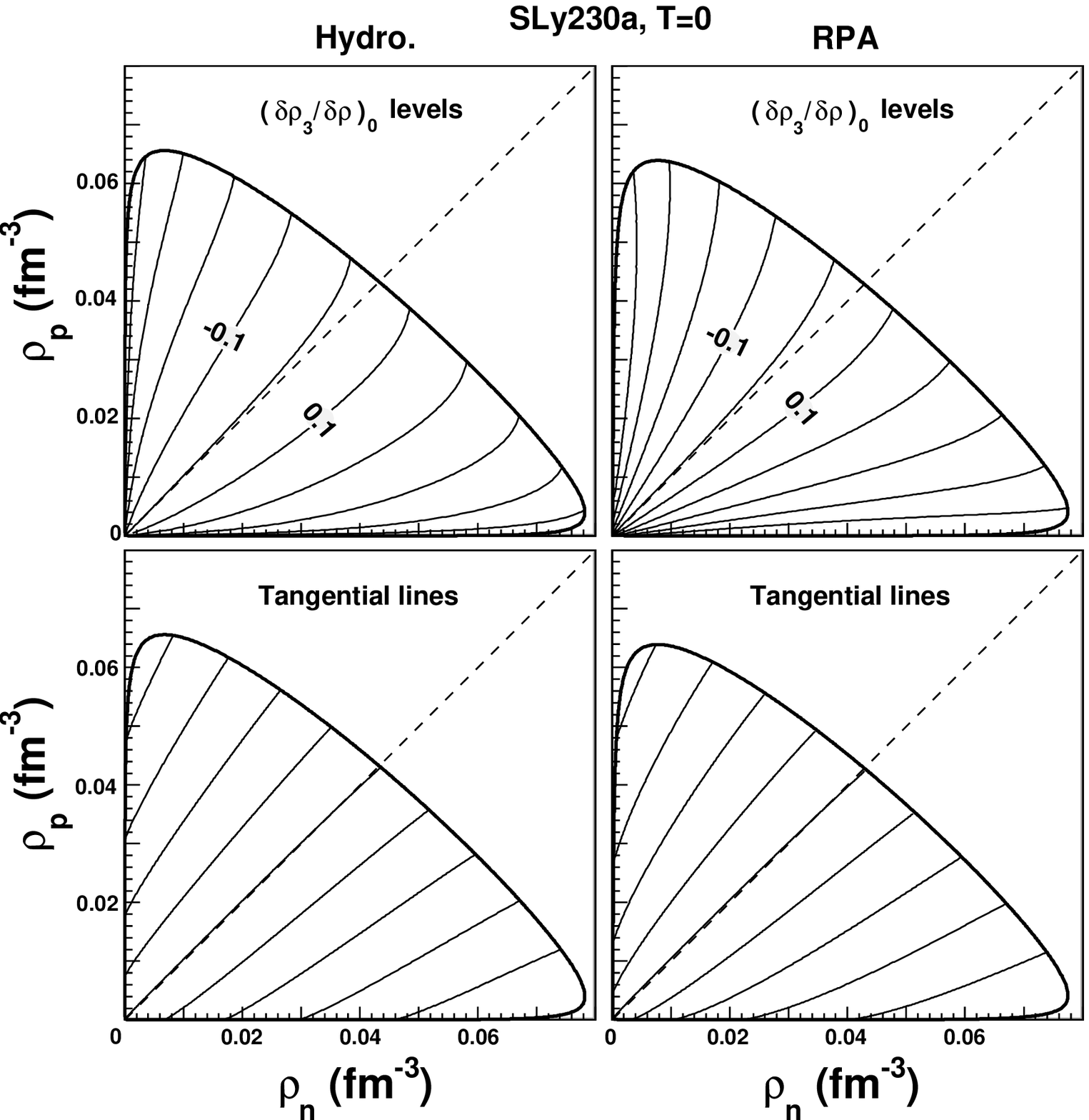}
\caption{(color online) 
	Phase-separation directions according to the most unstable mode, 
	reprented by levels and tangential lines 
	(see text for detailed explanation)
	inside the instability envelope (thick curve).   
	RPA and hydrodynamics results are presented. 
	Levels are separated by intervals of $0.1$.  
	}
\label{FIG:rnrp_dir0}
\end{center}
\end{figure}

We show on Fig.~\ref{FIG:rnrp_tau0l0} the projection on the density plane
of the iso-contours associated with the typical time constant $\tau_0$ (top) 
and fluctuation size $\lambda_0/2$ (bottom). 
Hydrodynamics and RPA results are shown, respectively on the left and right panels.
The  instability  envelope corresponds to  $\tau_0=\infty$.
Inside this envelope, the region of fastest time evolution corresponds to the lowest  values of $\tau_0$ 
and is found for low-density symmetric matter ($\rho \simeq 0.03$~fm$^{-3}$ for both approaches).
The RPA approach has typical times twice larger than the hydrodynamical ones.
The half-wavelength $\lambda_0/2$ is larger in the quantal approach than in the semi-classical one.
These results generalize the features observed 
on the dispersion relation $1/\tau(\q)$ illustrated on Fig.~\ref{FIG:disp},
\emph{i.e.} the quantum quenching of instabilities and the higher instability in $\q$ and $1/\tau$ of the first 
sound compared to zero sound.

Let us stress here that, although larger time constants and larger cluster sizes are repeatedly associated
in the comparison between the different approaches, 
these two features are actually decoupled.
Indeed, $\tau_0$ and $\lambda_0$ are not linked by a proportionality relation,
since they are determined by the minimum-$\tau$ relation $({\rm d}\tau/{\rm d}\q)_0=0$. 
In fact, the observed correlation is due to the particular shapes of the respective
dispersion curves $1/\tau(\q)$.
It can even be observed on the low-density part of the hydrodynamics plots 
that longer times can be associated with shorter wavelengths,
stressing the absence of proportionality relation between those two quantities.

It is interesting to give an estimate of the cluster masses associated with the favored wavelengths 
represented on the bottom part of Fig.~\ref{FIG:rnrp_tau0l0}. 
Let us first notice that the precise determination of this quantity is beyond the scope of the present
approach, which describes the initial dynamics of fragment formation; 
the properties of the fully separated fragments in the final state coming from the spinodal decomposition should be obtained following the evolution of the density fluctuation beyond the linear regime. 
However, an order of magnitude can be obtained in the present framework 
assuming that the half-wavelength $\lambda_0/2$ of the favored mode
could give an estimate of the nucleus diameter at saturation density.
In Tab.~\ref{tab:cluster-size-estimate} are listed some values of the half-wavelength $\lambda_0/2$ 
of the favored mode and the deduced mass of produced nuclei.
Comparing the typical values of the half-wavelength $\lambda_0/2$ shown on Fig.~\ref{FIG:rnrp_tau0l0} with the masses obtained in Tab.~\ref{tab:cluster-size-estimate}, 
it could be inferred that the spinodal decomposition scenario 
predicts the formation of light nuclei (up to A$\simeq 20$) through most of the unstable region.
Medium-mass nuclei could be formed in the region close to the frontiers of instability
(large global density or neutron excess):
their formation would thus require much more time than that of lighter elements.

\begin{table}[t]
\begin{center}
\renewcommand{\arraystretch}{1.5}
\begin{tabular}{|c|c|c|c|c|c|c|c|c|c|c|c|c|c|}
\hline 
$\lambda_0/2$ [fm] & $4$ & $5$ & $6$ & $7$ & $8$ & $9$ & $10$ & $11$ & $12$ & $13$ & $14$ & $15$ \\
\hline 
A & $5$ & $9$ & $16$ & $25$ & $37$& $53$ & $72$ & $96$ &$125$ & $159$ & $199$ & $244$ \\
\hline 
masses & \multicolumn{3}{c|}{light}&\multicolumn{5}{c|}{medium}& \multicolumn{4}{c|}{heavy}\\
\hline
\end{tabular}
\end{center}
\caption{
Relation between the half-wavelength of the favored mode and the
mass-number $A$ of a nucleus at normal density,  
$\lambda_0/2=2r_0A^{1/3}$, with $r_0=1.2$~fm.
}
\label{tab:cluster-size-estimate}
\end{table}%

On Fig.~\ref{FIG:rnrp_dir0}, we explore the instability direction $(\delta\rho_3/\delta\rho)_0$
through the density plane, for RPA and hydrodynamics.
The top view-graphs display level-lines of the quantity $(\delta\rho_3/\delta\rho)_0$.
A different kind of representation is employed for the bottom view-graphs:
they give a direct picture of the instability direction in the plane,
by the so-called tangential lines. 
In each point of a tangential line, the tangent to this curve gives the direction of the most unstable mode.
This representation clearly shows the isoscalar character of the modes obtained in both approaches.

Let us now comment the level-line panels of Fig.~\ref{FIG:rnrp_dir0}.
The $(\delta\rho_3/\delta\rho)_0$ level corresponding to the value $0$ is the purely isoscalar mode.
In the absence of Coulomb interaction, 
this level strictly corresponds to the axis of symmetric matter~\cite{bar06},
as imposed by the isospin-exchange symmetry.
In the present case, we have already noticed that the Coulomb interaction disfavors proton-density fluctuations.
This results in a global increase of the $(\delta\rho_3/\delta\rho)_0$ values,
thus the levels are shifted towards the proton-rich side.
This effect also appears on the left part of Fig.~\ref{FIG:qcte}:
as expected, the Coulomb shift is increased for larger values of $\q$ (middle graph).
Coming back to Fig.~\ref{FIG:rnrp_dir0},
let us compare the hydrodynamics and RPA upper panels. 
Two differences can be noticed:
i) in RPA the absolute values of $(\delta\rho_3/\delta\rho)_0$ are larger,
and ii) level-lines on the neutron-rich side approximately follow constant-$Z/A$ lines
while hydrodynamics levels are incurved. 
These panels thus give a global view of the features 
which have already appeared on Figs.~\ref{FIG:disp} and~\ref{FIG:qcte},
namely a reduced distillation in RPA, 
with a weaker density-dependence of $(\delta\rho_3/\delta\rho)_0$ at fixed $Z/A$.

\begin{figure}[t]
\begin{center}
\includegraphics[width=0.9\linewidth]{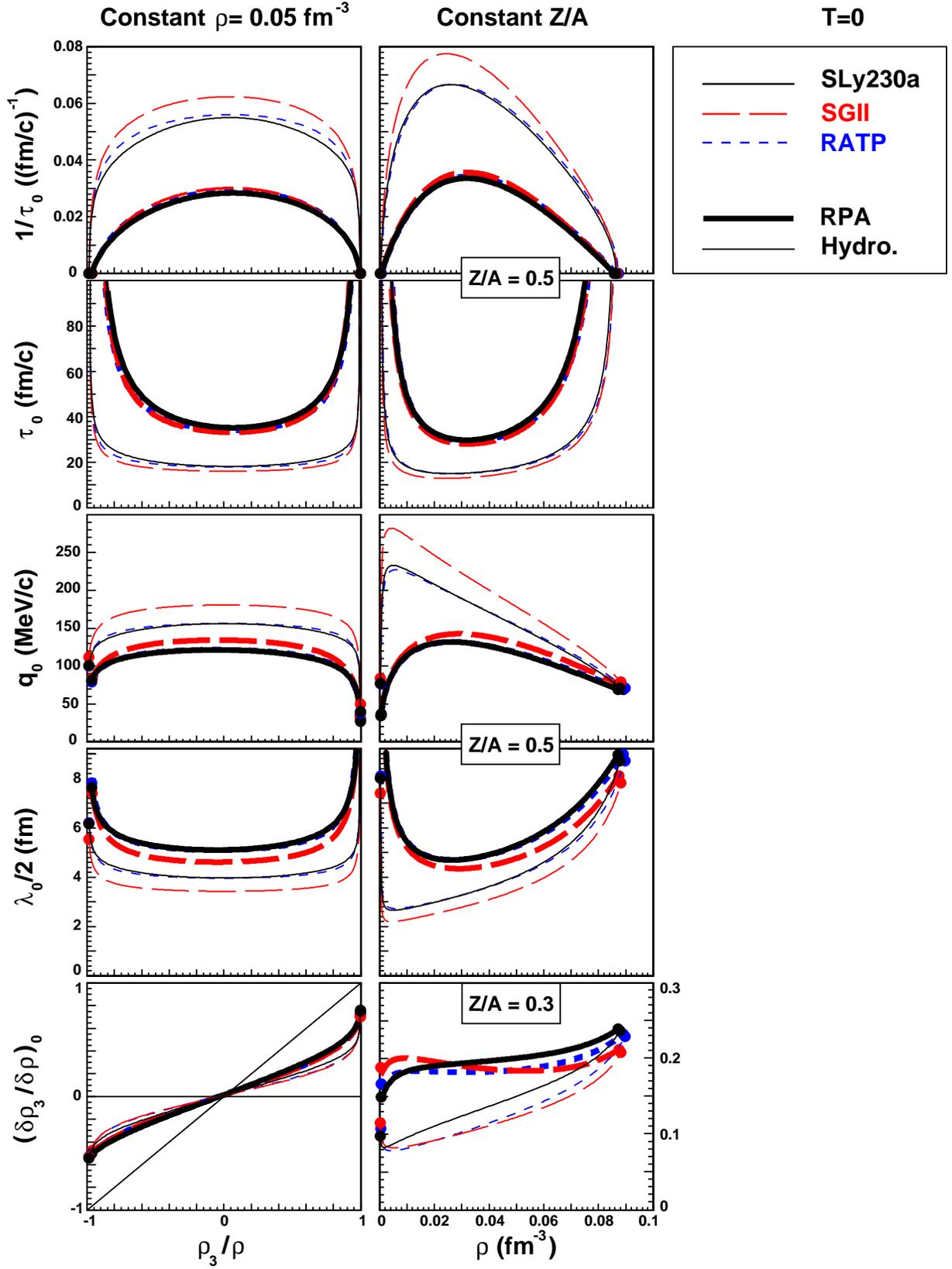}
\caption{(color online) 
	Time, size and isospin fractionation characterizing the most unstable mode:
	dependence on the Skyrme parametrization.
	RPA (thick line) and hydrodynamics (thin line) results are shown.
	Left: constant density.
	Right: constant proton fraction.	
	}
\label{FIG:mpi_dep_force}
\end{center}
\end{figure}

\begin{figure}[t]
\begin{center}
\includegraphics[width=0.9\linewidth]{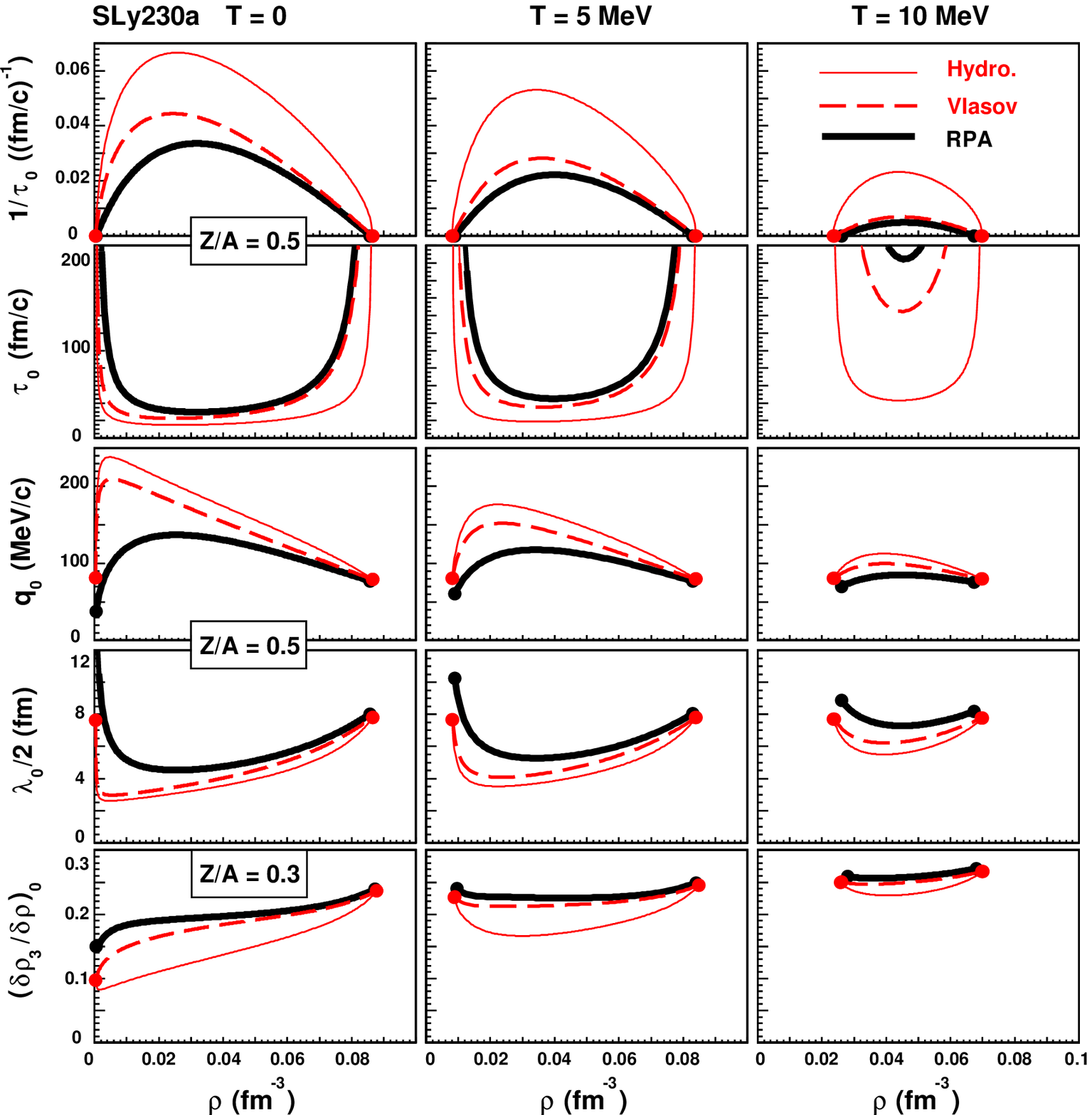}
\caption{(color online) 
	Time, size and isospin fractionation characterizing the most unstable mode:
	temperature evolution.
	RPA (thick line), Vlasov (dashed line) and hydrodynamics (thin line) results are shown.
	Constant proton fraction: 
	$Z/A=0.5$ for time and size characteristics (four upper lines),
	$Z/A=0.3$ for isospin fractionation.
	}
\label{FIG:mpi_dep_T}
\end{center}
\end{figure}

Figure~\ref{FIG:mpi_dep_force} displays the clusterization properties 
obtained along selected axis, for different Skyrme forces, at $T=0$.
Time, size and direction characterizing the most unstable mode are addressed,
comparing RPA and hydrodynamics results.
The inverse typical times $1/\tau_0$ differ typically by a factor of $2$ between the two approaches.
The cluster sizes are generally increased by about two fermis in the RPA approach.
At very low density however, much shorter wavelength are favored in the hydrodynamical approach:
note that the validity of the Thomas-Fermi approximation is under question in this domain.
Finally, the curves giving the direction of the most-unstable mode are very similar to
those obtained on Fig.~\ref{FIG:qcte} for $\q=100$~MeV/c:
this is due to the weak dependence of the instability direction
$\delta\rho_3/\delta\rho$ on the momentum $\q$ for large enough $\q$ values
(see Fig.~\ref{FIG:disp}).
Let us now compare the different Skyrme forces. 
Similar results are obtained with the three parametrizations.
SGII presents faster time evolutions and favors shorter cluster sizes,
as was already visible on Fig.~\ref{FIG:disp}. As for the direction of phase separation,
we can notice that the isospin-distillation effect is slightly weaker with SLy230a.

Finally, we explore on Fig.~\ref{FIG:mpi_dep_T} the temperature dependence of the clusterization properties.
Time, size and direction characteristics are shown for $T=0$, $5$ and $10$~MeV,
comparing now the three approaches: RPA, Vlasov and hydrodynamics.
The differences between RPA and Vlasov results give an estimate of the quantum effects.
As can be seen from the size characteristics, quantum effects are still sensible at $T=10$~MeV:
the Vlasov curve for $\q_0$ is clearly distinct from the RPA one.
The smoothing of the differences of clusterization properties 
between the three approaches with increasing temperature
is mainly due to the reduction of the instability.
The range of $\q$ and $\delta\rho_3/\delta\rho$ values
corresponding to a negative free-energy curvature is narrowed, 
constraining the different approaches to a closer agreement
as for the size and direction associated with the most unstable mode.

The time-evolution part of Fig.~\ref{FIG:mpi_dep_T} deserves a special comment. 
As $T$ increases, all values of $1/\tau_0$ go towards zero, 
which defines the limit of the instability region.
At $T=10$~MeV, Vlasov and RPA results for this quantity are quite close:
however, if we still increase the temperature, the RPA curve will disappear first,
then Vlasov and hydrodynamics curves will reach zero together. 
While the $1/\tau_0$ values converge to zero,
the corresponding divergence of $\tau_0$ on the contrary 
magnifies the differences between the three approaches.
As a result, the range of prediction as for the time-scale of cluster formation
is strongly increased as we approach the limit of the instability region.

\section{Conclusion}
\label{SEC:conclusion}

We have presented the clusterization properties of asymmetric nuclear matter in three approaches,
comparing the quantal RPA results
with those obtains in a semi-classical framework, with Vlasov and hydrodynamical calculations.
In all cases, starting from homogeneous infinite matter,
we have studied the consequences of introducing 
a plane-wave density fluctuation of finite momentum $\q$,
working in the small-amplitude limit.
Liquid-gas instabilities have thus been identified, corresponding to an amplification of the fluctuation.
The unstable eigen-modes (defined for each approach)
are characterized by a time constant $\tau$, a wavelength $\lambda=2\pi/\q$,
and a direction $\delta\rho_3/\delta\rho$.
Assuming a scenario of spinodal decomposition, 
the early dynamics of cluster formation is dominated by the most unstable mode
(for given temperature and densities).
The clusterization properties are then characterized by the quantities associated with this mode:
a time-scale $\tau_0$, a typical cluster size $\lambda_0/2=\pi/\q_0$,
and a cluster composition induced by the direction $(\delta\rho_3/\delta\rho)_0$.
We have noticed that this study, which assumes fragment formation out of equilibrium,
is related to the instability properties of homogeneous matter at sub-saturation density;
it is complementary to a description of equilibrated, clusterized matter.

Considering infinite nuclear matter is an approximation 
which permits to treat analytically in a common framework
the liquid-gas instabilities concerning both compact stars 
and the nuclear multifragmentation observed in heavy-ion collisions.
In this framework, the differences between the presented approaches
can be attributed to two well-identified origins:
comparing Vlasov and RPA gives a direct estimate of the quantum effects,
while comparing Vlasov and hydrodynamics 
shows the difference between a collisionless and local equilibrium description.

The results obtained with the different approaches show similar qualitative features, 
reproduced with the three Skyrme-like forces that we have used (SLy230a, SGII, RATP).
In particular, the regions of instability against finite-size density fluctuations are very close,
and the direction of phase separation is always found to be essentially isoscalar,
favoring the phenomenon of isospin distillation which leads to more symmetric clusters.
An order of magnitude of the cluster masses, estimated as spherical nuclei of diameter $\lambda_0/2$, indicates that light to medium nuclei ($A<40$) should be favored through most of the instability region for the various approaches.
However, some quantitative differences are noticeable.
As for the comparison between the Skyrme forces,
the main distinction concerns SGII, associated with shorter size and time characteristics.
Let us now summarize the observations concerning the three different approaches.
The comparison between RPA and Vlasov
shows a quantum quenching leading to larger size and time characteristics in the RPA framework.
This effect survives for all the temperature range of the instability.
The Vlasov results are framed between the RPA and hydrodynamics ones,
which can be considered as giving two opposite limits of predictions.
The physical processes should correspond to an intermediate situation,
involving different possible relaxation rates.
This is an important aspect to take into account in realistic calculations
aiming at a quantitative description of nuclear fragmentation.
Indeed the two limits we have obtained in the present work 
indicate that the interval of predictions can be quite large.
At $T=0$, the inverse time constant $1/\tau_0$ is typically enhanced by a factor $2$
between RPA (slower) and hydrodynamics (faster).
The cluster size is about $2$ fermis smaller for the hydrodynamics,
with a stronger reduction at low density 
where the validity of the semi-classical approaches gets more questionable (due to the low Fermi momenta).
The direction of the unstable modes, for a given value of $\q$, is little sensitive to the quantum effects.
The main difference is observed between collisionless and local equilibrium approaches:
inside the instability region, the direction corresponding to a divergence of the response function
is slightly less isoscalar than the direction of minimal free-energy curvature.

Approaching the frontiers of the instability region (in densities or temperature),
the predictions between all approaches get closer
as for the size and direction induced by the most unstable mode,
due to a reduction of the instability interval of $\q$ and $\delta\rho_3/\delta\rho$ values.
The time constant $\tau_0$ characterizing the rapidity of cluster formation
has a specific behavior, since it diverges (by definition) on the borders of the instability region:
the range of values obtained with the different approaches is then strongly enhanced.
This effect is important to notice, since in heavy-ion collisions 
as well as core-collapse supernovae
the time-scale of cluster formation should be compared with that of various equilibration processes.
Conversely, a deeper study of the different time scales involved
could indicate the respective validity of collisionless and local equilibrium descriptions of matter clusterization.


\appendix

\section{Linear response to an external probe}
\label{APP:response}

We detail the calculation of the response function for  the single-fluid case.
Let us consider the response to an infinitesimal external field of the form:
\begin{equation}
\begin{array}{llll}
\hat{\mathrm{V}}_{\rm ext} = \mathcal E {\rm e}^{{\rm i}\qv\cdot \hat{\rv}-{\rm i}(\omega+{\rm i}\eta) t} & (t<0) & ; \; 
\hat{\mathrm{V}}_{\rm ext} = 0 & (t \geq 0)
\;,
\end{array}
\end{equation}
carrying a momentum $\qv$ and an energy $\omega$
(the positive infinitesimal part $\eta$ ensuring the adiabatic switching of the external field).
At late enough times and small enough external field $\mathcal E$, the expectation
value of the density fluctuation has the same space and time
dependence as the external field:
\be
\delta\rho(\rv,t)&=&\alpha {\rm e}^{{\rm i}\qv\cdot\rv-{\rm i}(\omega+{\rm i}\eta) t}=\tilde{\alpha}(t){\rm e}^{{\rm i}\qv\cdot\rv} \;,
\ee
where $\tilde{\alpha}$ includes the time dependence for a more compact notation.
The linear response $\Pi(\omega,\qv)$ is defined by the
ratio of the density change to the external field strength,
for a small external field $\mE$:
\begin{equation}
\Pi(\omega,\qv) = 
\frac{\alpha}{\mE} \;.
\end{equation}
A density fluctuation $\delta \hat \rho (t)$ defined such as
$\hat \rho (t)= \hat \rho_0 + \delta \hat \rho (t)$
induces a variation $\delta \hat{\mathrm{h}}(t)$ of the mean-field operator, such that
$\hat{\mathrm{h}}(t)= \hat{\mathrm{h}}_0 + \delta \hat{\mathrm{h}} (t)$.
The linearized TDHF equation
\begin{equation}
\label{EQ-an:TDHF-lin}
i\hbar \frac{\partial(\delta \hat \rho)}{\partial t} 
=  \left[ \hat{\mathrm{h}}_0,\delta \hat \rho \right] 
+ \left[\delta \hat{\mathrm{h}} + \hat{\mathrm{V}}_{\rm ext},\hat \rho_0 \right]
\end{equation}
reads in momentum space:
\be
\label{EQ-an:TDHF-lin-element}
\langle\kv| \delta \hat \rho |\kpv\rangle 
= \frac{g(\mathrm{f}_{\kpv}-\mathrm{f}_{\kv})}{\hbar[\omega-(\omega_{\kv}-\omega_{\kpv})+{\rm i}\eta]} 
\langle\kv| \delta \hat{\mathrm{h}} + \hat{\mathrm{V}}_{\rm ext} |\kpv\rangle \;.
\ee
The elements of $\hat{\mathrm{V}}_{\rm ext}$  are:
\be
\langle\kv| \hat{\mathrm{V}}_{\rm ext} |\kpv\rangle 	
& = & \mE {\rm e}^{-{\rm i}(\omega+{\rm i}\eta)t} \delta_{\kpv+\qv,\kv} = \tilde{\mE}(t)\; \delta_{\kpv+\qv,\kv}
\ee
and the elements of $\delta\hat\rho$ have the form:
\be
\label{EQ-an:drho-element}
\langle\kv|\delta\hat\rho|\kpv\rangle 
&=& \kappa(\kv,\qv){\rm e}^{-{\rm i}(\omega+{\rm i}\eta)t} \delta_{\kpv+\qv,\kv}
= \tilde{\kappa}(\kv,\qv,t) \; \delta_{\kpv+\qv,\kv} \;.
\ee
The density factors $\alpha$ and $\kappa(\kv,\qv)$ are related by:
\be
\alpha 
&=& \frac{1}{\Omega}\sum_{\kv} \kappa(\kv,\qv)
\;.
\ee
According to Eqs.~(\ref{EQ-an:TDHF-lin-element}-\ref{EQ-an:drho-element}), 
the matrix elements of $\delta\hat{\mathrm{h}}$ are of the form:
\be
\label{EQ-an:dh-element-1}
\langle\kv|\delta\hat{\mathrm{h}}|\kpv\rangle 
&=& \nu(\kv,\qv)  {\rm e}^{-{\rm i}(\omega +{\rm i}\eta)t} \delta_{\kpv+\qv,\kv} 
= \tilde{\nu}(\kv,\qv,t)\; \delta_{\kpv+\qv,\kv} \;,
\ee
where $\nu(\kv,\qv)$ is connected to the residual interaction $\hat{\mathrm{V}}^{\rm res}$. 
Denoting $\mathrm{E}[\hat\rho]$ the total energy of the system, 
this residual interaction operator is defined by~\cite{Ring-Schuck}:
\be
\delta \mathrm{E}
=\mathrm{E}[\hat\rho_0+\delta\hat\rho]-\mathrm{E}[\hat\rho_0]
&=&\sum_{ij}\frac{\partial \mathrm{E}}{\partial\rho_{ij}} \delta\rho_{ij}
+ \frac{1}{2}\sum_{ijkl}\frac{\partial^2 \mathrm{E}}{\partial\rho_{kl}\partial\rho_{ij}}\delta\rho_{ij}\delta\rho_{kl}\nonumber\\
&=&\sum_{ij} \mathrm{h}_{ji}\delta\rho_{ij}+\frac{1}{2}\sum_{ijkl}\mathrm{V}^{\rm res}_{jl;ik}\delta\rho_{ij}\delta\rho_{kl}\nonumber\\
\label{EQ-an:dE-micro-DFT}
&=&{\rm Tr} \left[\hat{\mathrm{h}} \delta\hat\rho\right] 
+ \frac{1}{2}{\rm Tr}_{(1)}{\rm Tr}_{(2)} \left[ \hat{\mathrm{V}}^{\rm res}_{(1,2)}\delta\hat\rho_{(1)}\delta\hat\rho_{(2)} \right] 
\ee
where we have limited the development to the second order in $\delta\rho_{ij}$.
From these relations, we also have the expression of the 
mean-field variation in terms of the residual interaction (at first order in $\delta\rho_{ij}$):
\be
\delta \mathrm{h}_{ji} 
&=& \sum_{kl} \frac{\partial^2 \mathrm{E}}{\partial \rho_{kl}\partial\rho_{ij}}\delta\rho_{kl}
=\sum_{kl} \mathrm{V}^{\rm res}_{jl;ik}\delta\rho_{kl}\\
\delta\hat{\mathrm{h}} &=& {\rm Tr}_{(2)} \left[\hat{\mathrm{V}}^{\rm res}_{(1,2)}\delta\hat\rho_{(2)}\right]
\;,
\ee
so that from Eq.~(\ref{EQ-an:dh-element-1}) we have:
\be
\langle\kv+\qv|\delta\hat{\mathrm{h}}|\kv\rangle
=\tilde{\nu}(\kv,\qv,t)
&=& \sum_{\kv_2\kpv_2} 
\mathrm{V}^{\rm res}_{\kv+\qv,\kv_2 \:;\: \kv,\kpv_2} \; \langle\kpv_2|\delta\hat \rho|\kv_2\rangle
\nonumber\\
&=& \sum_{\kv_2} 
\mathrm{V}^{\rm res}_{\kv+\qv,\kv_2 \:;\: \kv,\kv_2+\qv} \; \tilde{\kappa}(\kv_2,\qv,t)
\;.
\ee
Assuming that the residual interaction element 
$\mathrm{V}^{\rm res}_{\kv+\qv,\kv_2 \:;\: \kv,\kv_2+\qv}$ does not
depend on $(\kv,\kv_2)$ - Landau monopolar approximation - it can be
written: 
\be
\mathrm{V}^{\rm res}_{\kv+\qv,\kv_2 \:;\: \kv,\kv_2+\qv} &=& \mathrm{V}^{\rm res}(\qv)
\ee
and we obtain:
\be
\tilde{\nu}(\kv,\qv,t) = \tilde{\nu}(\qv,t) 
&=& \mathrm{V}^{\rm res}(\qv) \sum_{\kv} \tilde{\kappa}(\kv,\qv,t) 
= \Omega \mathrm{V}^{\rm res}(\qv)\tilde{\alpha}(t)
= \mathrm{v}^{\rm res}(\qv)\tilde{\alpha}(t)
\ee
with the notation $\mathrm{v}^{\rm res}=\Omega \mathrm{V}^{\rm res}$.
The matrix elements of $\delta\hat{\mathrm{h}}$ are then:
\be
\label{EQ-an:dh-element-2}
\langle\kv|\delta\hat{\mathrm{h}}|\kpv\rangle 
&=& \mathrm{v}^{\rm res}(\qv)\tilde{\alpha}(t) \; \delta_{\kpv+\qv,\kv} \;.
\ee

Inserting the expressions obtained for the matrix elements of 
$\delta\hat\rho$, $\delta\hat{\mathrm{h}}$ and $\hat{\mathrm{V}}_{\rm ext}$,
the linearized TDHF equation~(\ref{EQ-an:TDHF-lin-element}) now reads:
\be
\tilde{\kappa}(\kv,\qv,t)\delta_{\kpv+\qv,\kv}
&=&\frac{g(\mathrm{f}_{\kpv}-\mathrm{f}_{\kv})}{\hbar[\omega-(\omega_{\kv}-\omega_{\kpv})+{\rm i}\eta]} 
\left(\mathrm{v}^{\rm res}\tilde{\alpha}(t)+\tilde{\mE}\right)\delta_{\kpv+\qv,\kv}  \\
\label{EQ-an:reponse-discrete}
\sum_{\kv}\tilde{\kappa}(\kv,\qv,t) = \Omega\tilde{\alpha}(t)
&=&
\sum_{\kv} \frac{g(\mathrm{f}_{\kv}-\mathrm{f}_{\kv+\qv})}{\hbar[(\omega-\omega_{\mathbf{kq}})+{\rm i}\eta]}
\left(\mathrm{v}^{\rm res}\tilde{\alpha}(t)+\tilde{\mE}\right)
\ee
where $\omega_{\mathbf{kq}}=\omega_{\kv+\qv}-\omega_{\kv}$. The second line is obtained by summing over $\kv$ and $\kpv$. 
Taking the continuous limit of Eq.~(\ref{EQ-an:reponse-discrete}) and multiplying by the time factor 
${\rm e}^{-{\rm i}(\omega-{\rm i}\eta)t}$, we obtain:
\be
\label{EQ-an:reponse-continue}
\alpha&=&\int{\frac{{\rm d}\kv}{(2\pi)^3} 
\frac{g(\mathrm{f}_{\kv}-\mathrm{f}_{\kv+\qv})}{\hbar[(\omega-\omega_{\mathbf{kq}})+{\rm i}\eta] }
\left( \mathrm{v}^{\rm res}\alpha+\mE \right)}
\nonumber\\
\label{EQ-an:reponse-continue-Pi0}
&=& \Pi_0 (\omega,\qv) \left( \mathrm{v}^{\rm res}\alpha+\mE \right)
\;.
\ee
where $\Pi_0$ is the uncorrelated response function (Lindhard function).
A rearrangement of Eq.~(\ref{EQ-an:reponse-continue-Pi0}) gives the following expression
for the response function:
\be
\label{EQ-an:polar}
\Pi(\omega,\qv)= \frac{\alpha}{\mathcal E} = \frac{\Pi_0(\omega,\qv)}{1-\Pi_0(\omega,\qv) \mathrm{v}^{\rm res}} \;.
\ee
This result could also be deduced from the Bethe-Salpeter equation~\cite{Fetter-Walecka}, 
keeping the transfered-momentum dependence of the residual approximation 
and treating the exchange terms at the Landau approximation:
this is the LAFET approximation detailed in Ref.~\cite{mar05}.

\section{Residual interaction and free-energy curvature}
\label{APP:residual}

In order to establish the link between the residual interaction and the free-energy curvature,
we start with the thermodynamic case considering a homogeneous system 
with large particle number $N$ and volume $\Omega$. 
For simplicity we first expose the calculation details in the single-fluid case.
Fixing $\Omega$, at equilibrium the density matrix $\hat\rho$ is uniquely defined by $N$
and a fluctuation of the particle number $\delta N$ leads to a fluctuation $\delta\hat\rho$
which can be expressed by the Taylor expansion:
\be
\delta\hat\rho
&=&\frac{\partial\hat\rho}{\partial \mathrm{N}}\delta \mathrm{N} 
+\frac{1}{2}\frac{\partial^2\hat\rho}{\partial \mathrm{N}^2}\delta \mathrm{N}^2+o(\delta \mathrm{N}^3)\;,
\ee
and the free-energy variation is:
\be
\delta \mathrm{F}^{\rm b} 
&=&\delta \mathrm{E}^{\rm b}-T\delta \mathrm{S}\nonumber\\
&=& \mu\delta \mathrm{N} + \frac{1}{2}\frac{\partial\mu}{\partial \mathrm{N}}\delta \mathrm{N}^2 +o(\delta \mathrm{N}^3)
\label{EQ-an:dFb-1}
\ee
where $\mathrm{S}$ is the entropy and $\mu$ the chemical potential.
The superscripts "$\rm b$" (for "bulk") specify that we are in the thermodynamic description.
$\delta \mathrm{F}^{\rm b}$ can be separated into an non-interacting
term $\delta \mathrm{F}^{\rm b,NI}$ 
and a residual interaction term $\delta \mathrm{F}^{\rm b,res}$.
The non-interacting part gives the free-energy variation for non-interacting particles
of mass $\mathrm{m}^*$ in a constant mean field. It contains the entropy variation, 
since the entropy $\mathrm{S}$ is determined by the density
independently of the particle interaction. 
The residual interaction term then reduces to 
$\delta \mathrm{F}^{\rm b,res}=\delta \mathrm{E}^{\rm b,res}$, 
which is the term due to the variation of the mean field 
$\delta\hat{\mathrm{h}}$, so that:
\be
\delta \mathrm{F}^{\rm b} 
&=& \delta \mathrm{F}^{\rm b,NI}+\delta \mathrm{E}^{\rm b,res} 
\;.
\ee
Referring to the expression (\ref{EQ-an:dE-micro-DFT}),
$\delta \mathrm{E}^{\rm b,res}$ is expressed in terms of the residual-interaction operator as:
\be
\delta \mathrm{E}^{\rm b,res}
&=&\frac{1}{2}{\rm Tr}_{(1)}{\rm Tr}_{(2)} \left[ \hat{\mathrm{V}}^{\rm b,res}_{(1,2)}\delta\hat\rho_{(1)}\delta\hat\rho_{(2)} \right]\nonumber\\
&=&\frac{1}{2}{\rm Tr}_{(1)}{\rm Tr}_{(2)} \left[ \hat{\mathrm{V}}^{\rm b,res}_{(1,2)}
\frac{\partial\hat\rho_{(1)}}{\partial \mathrm{N}}\frac{\partial\hat\rho_{(2)}}{\partial \mathrm{N}}\right]\delta \mathrm{N}^2 
+ o(\delta \mathrm{N}^3)
\ee
Taking constant matrix elements for $\hat{\mathrm{V}}^{\rm b,res}$,
such that $\mathrm{V}^{\rm b,res}_{jl;ik}=\mathrm{V}^{\rm b,res}$, 
we obtain (\emph{cf} Appendix \ref{APP:calc-vres}):
\be
\label{EQ-an:dE-b-res-scal}
\delta \mathrm{E}^{\rm b,res}&=&\frac{1}{2}\mathrm{V}^{\rm b,res}\delta \mathrm{N}^2+o(\delta \mathrm{N}^3) \;,
\ee
so that
\be
\label{EQ-an:dFb-2}
\delta \mathrm{F}^{\rm b}&=&\mu\delta \mathrm{N} + 
\frac{1}{2}\left[ \left(\frac{\partial\mu}{\partial \mathrm{N}}\right)^{\rm NI}
  + \mathrm{V}^{\rm b,res} \right]\delta \mathrm{N}^2
+o(\delta \mathrm{N}^3)
\;.
\ee
Identifying the second-order term of Eqs.~(\ref{EQ-an:dFb-1}) and (\ref{EQ-an:dFb-2}) gives:
\be
\label{EQ-an:id-ordre2-1}
\frac{\partial\mu}{\partial \mathrm{N}}&=&\left(\frac{\partial\mu}{\partial
  \mathrm{N}}\right)^{\rm NI} + \mathrm{V}^{\rm b,res}
\;.
\ee
The non-interacting part of the chemical-potential derivative is
related to the density of states at the Fermi level, denoted
$\mathrm{N}_0$. Taking the derivative with respect to the density
$\rho=N/\Omega$, 
we have:
\be
\left(\frac{\partial \mu}{\partial \rho}\right)^{\rm NI}
&=&\frac{\hbar^2}{2\mathrm{m}^*}\frac{\partial k_{\rm F}^2}{\partial\rho} 
= \frac{1}{\mathrm{N}_0}
\;,
\ee
where $k_{\rm F}$ is the Fermi momentum.
Using the notation $\Omega \mathrm{V}^{\rm b,res}=\mathrm{v}^{\rm b,res}$, 
we can re-express Eq.~(\ref{EQ-an:id-ordre2-1}) as: 
\be
\label{EQ-an:id-ordre2-2}
\frac{\partial\mu}{\partial\rho}&=&\frac{1}{\mathrm{N}_0}+ \mathrm{v}^{\rm b,res}
\;,
\ee
which establishes the link between the thermodynamic free-energy curvature
and the bulk part of the residual interaction.

Let us now consider a finite-size density fluctuation.
In order to obtain the expression of the residual interaction 
in terms of the free-energy curvature matrix, we still work in the local equilibrium approach.
Following Eq.~(\ref{EQ-an:dE-b-res-scal}), the bulk part of the residual energy variation per unit volume reads (dropping the terms beyond second order in $\delta\rho$):
\be
\label{EQ-an:dEbres-div-Omega}
\frac{\delta \mathrm{E}^{\rm b,res}}{\Omega}
&=&\frac{1}{2}\Omega \mathrm{V}^{\rm b,res} \delta\rho^2 
= \frac{1}{2} \mathrm{v}^{\rm b,res} \delta\rho^2 
\;.
\ee
Considering a density fluctuation $\delta\rho(\rv)$, 
the average residual energy variation per unit volume is (for the bulk part):
\be
\delta\mE^{\rm b,res}
&=&\frac{1}{\Omega}\int{{\rm d}\rv\; \frac{\delta \mathrm{E}^{\rm b,res}[\delta\rho(\rv)]}{\Omega}}
=\frac{1}{2\Omega}\int{{\rm d}\rv\; \mathrm{v}^{\rm b,res} \delta\rho(\rv)^2}
\ee
With a Fourier development of $\delta\rho(\rv)$:
\be
\delta\rho(\rv)=\sum_{\qv}\rho_{\qv}
&;&
\rho_{\qv}=\mathrm{A}_{\qv}{\rm e}^{{\rm i}\qv\cdot\rv}
\;\;\; ; \;\;\; \mathrm{A}_{-\qv}=\mathrm{A}^*_{\qv}
\;,
\ee
we have:
\be
\delta\mE^{\rm b,res}&=&\frac{1}{2}\sum_{\qv} |\mathrm{A}_{\qv}|^2 \mathrm{v}^{\rm b,res} \;.
\ee
Adding a transfered-momentum dependence to the residual interaction,
we generalize the previous expression to:
\be
\label{EQ-an:dmEres-dep-q}
\delta\mE^{\rm res}&=&\frac{1}{2}\sum_{\qv} |\mathrm{A}_{\qv}|^2 \mathrm{v}^{\rm res}(\qv)
\;.
\ee
The $\qv$-dependence concerns the gradient and Coulomb terms of the energy variation:
\be
\delta\mE^{\nabla}
&=&\frac{1}{\Omega}\int{{\rm d}\rv \mH^{\nabla}(\rv)}
=\frac{1}{\Omega}\int{{\rm d}\rv C^{\nabla}\left(\nabla\rho(\rv)\right)^2}
=\sum_{\qv} |\mathrm{A}_{\qv}|^2 \q^2 C^{\nabla}\\
\delta\mE^{\rm c}
&=&\frac{1}{2}\sum_{\qv} |\mathrm{A}_{\qv}|^2 \frac{4\pi e_i^2}{\q^2}
\ee
where $e_i=4\pi q_i^2/\epsilon_0$, $q_i$ being the particle electric charge
for the considered single-component fluid.
We can now express $\mathrm{v}^{\rm res}(\qv)$:
\be
\label{EQ-an:dmEres-dep-q-2}
\delta\mE^{\rm res}
&=&\delta\mE^{\rm b,res}+\delta\mE^{\nabla}+\delta\mE^{\rm c}
=\frac{1}{2}\sum_{\qv} |\mathrm{A}_{\qv}|^2 \left[ \mathrm{v}^{\rm b,res}+2C^{\nabla}\q^2+\frac{4\pi e_i^2}{\q^2} \right]\\
\mathrm{v}^{\rm res}(\qv)
&=&\mathrm{v}^{\rm b,res}+2C^{\nabla}\q^2+\frac{4\pi e^2}{\q^2}
=\left(\frac{\partial\mu}{\partial\rho}-\frac{1}{\mathrm{N}_0}\right)+2C^{\nabla}\q^2+\frac{4\pi e_i^2}{\q^2} \;.
\ee

The whole procedure can be easily generalized to the two-component nuclear matter,
defining the residual interaction operator by the relation (with $i,j={\rm n},{\rm p}$):
\be
\delta \mathrm{E}^{\rm res}
&=&\frac{1}{2}\sum_{i,j}\sum_{\kv_1\kv_2\kv_3\kv_4}
\frac{\partial^2\mathrm{E}}{\partial\rho_{i;\kv_1\kv_2}\partial\rho_{j;\kv_3\kv_4}}
\delta\rho_{i;\kv_1\kv_2}\delta\rho_{j;\kv_3\kv_4}\nonumber\\
&=&\frac{1}{2}\sum_{i,j} {\rm Tr}_{(1)}{\rm Tr}_{(2)}
\left[\hat{\mathrm{V}}^{\rm res}_{(i1,j2)}\delta\hat\rho_{(i1)}\delta\hat\rho_{(j2)}\right]
\;.
\ee
Equations
(\ref{EQ-an:id-ordre2-2}),
(\ref{EQ-an:dEbres-div-Omega}),
(\ref{EQ-an:dmEres-dep-q}) and
(\ref{EQ-an:dmEres-dep-q-2})
then become respectively:
\be
\frac{\partial\mu_i}{\partial\rho_j}&=&\mathrm{v}^{\rm b,res}_{ij}+\frac{\delta_{ij}}{\mathrm{N}_{0,i}}\\
\frac{\delta \mathrm{E}^{\rm b,res}}{\Omega}&=&\frac{1}{2}\sum_{i,j}\mathrm{v}^{\rm b,res}_{ij}\delta\rho_i\delta\rho_j\\
\delta\mE^{\rm res}
&=&\sum_{\qv} \sum_{i,j} \frac{\mathrm{A}_{i;\qv}\mathrm{A}^*_{j;\qv}}{2} \mathrm{v}^{\rm res}_{ij}(\qv)
\nonumber\\
&=&\sum_{\qv} \sum_{i,j} \frac{\mathrm{A}_{i;\qv}\mathrm{A}^*_{j;\qv}}{2}
\left[\mathrm{v}^{\rm b,res}_{ij}+2C^{\nabla}_{ij}\q^2+\frac{4\pi e_ie_j}{\q^2} \right]
\;,
\ee
and since 
\be
\mathrm{v}^{\rm res}_{ij}(\qv)=\mathrm{v}^{\rm b,res}_{ij}+2C^{\nabla}_{ij}\q^2+\frac{4\pi e_ie_j}{\q^2}
&=&\mathrm{C}^{\rm f}_{ij}(\qv) - \frac{\delta_{ij}}{\mathrm{N}_{0,i}} \;,
\ee
we obtain the following relation between the residual interaction and free-energy density curvature matrixes:
\be
\mathcal{C}^{\rm f} = \left( \mathbf{N_0} \right)^{-1} + \mathbf{v}^{\rm res} 
&;&
\mathbf{N_0} = \left(\begin{array}{cc}\mathrm{N}_{0,{\rm n}} & 0 \\0 & \mathrm{N}_{0,{\rm p}}
\end{array}\right) \; .
\ee
%

\section{Link between residual interaction $\hat{\mathrm{V}}^{\rm res}$ and energy variation $\delta \mathrm{E}^{\rm res}$}
\label{APP:calc-vres}

We first work in a discrete basis $|\kv_i\rangle$, where the residual energy variation is expressed by:
\be
2\delta \mathrm{E}^{\rm res}
&=&{\rm Tr}_{(1)}{\rm Tr}_{(2)}\left[\hat{\mathrm{V}}^{\rm res}_{(1,2)}\delta\hat\rho_{(1)}\delta\hat\rho_{(2)}\right]\nonumber\\
&=&\sum_{\kv_1\kpv_1}\sum_{\kv_2\kpv_2}
\langle\kpv_1\kpv_2|\hat{\mathrm{V}}^{\rm res}|\kv_1\kv_2\rangle
\langle\kv_1|\delta\hat\rho|\kpv_1\rangle
\langle\kv_2|\delta\hat\rho|\kpv_2\rangle
\ee
In the discrete basis, the one-body density matrix elements are:
\be
\langle\kv_i|\hat\rho|\kv_j\rangle=\mathrm{n}(\kv_i)\delta_{ij}
\;,
\ee
whose dependence on the particle number $\mathrm{N}$ is contained in the
occupation number $\mathrm{n}(\kv_i)=\mathrm{n}(\kv_i,\rho=\mathrm{N}/\Omega)$: 
\be
\langle\kv_i|\delta\hat\rho|\kv_j\rangle
&=&\langle\kv_i|\frac{\partial\hat\rho}{\partial \mathrm{N}}|\kv_j\rangle \delta \mathrm{N}
=\frac{1}{\Omega}\left[\frac{\partial \mathrm{n}(\kv_i,\rho)}{\partial \rho}\delta_{ij}\right]\delta \mathrm{N}
\;.
\ee
We have then for the residual interaction term:
\be
2\frac{\delta \mathrm{E}^{\rm res}}{\delta \mathrm{N}^2}
&=& \frac{1}{\Omega^2}  \sum_{\kv_1\kv_2} \langle\kv_1\kv_2|\hat{\mathrm{V}}^{\rm res}|\kv_1\kv_2\rangle \frac{\partial
  \mathrm{n}(\kv_1,\rho)}{\partial \rho} \frac{\partial \mathrm{n}(\kv_2,\rho)}{\partial
  \rho} 
\;. 
\ee

Let us now take the continuous limit.
Denoting $\langle\kv_1\kv_2|\hat{\mathrm{V}}^{\rm res}|\kv_1\kv_2\rangle=\mathrm{V}^{\rm res}_{\kv_1,\kv_2;\kv_1,\kv_2}$, 
the residual interaction term can be expressed as:
\be
\label{EQ-an:dEres-integ}
2\frac{\delta \mathrm{E}^{\rm res}}{\delta \mathrm{N}^2}
&=&\frac{1}{(2\pi)^6}\int{{\rm d}\kv_1 {\rm d}\kv_2 \mathrm{V}^{\rm
    res}_{\kv_1,\kv_2;\kv_1,\kv_2} \frac{\partial
    \mathrm{n}(\kv_1,\rho)}{\partial \rho} \frac{\partial
    \mathrm{n}(\kv_2,\rho)}{\partial\rho}} 
\;.
\ee
We now have to explicit $\mathrm{n}(\kv,\rho)$. 
For a fermion gas at thermodynamic equilibrium, at temperature $T=1/\beta$, 
with the chemical potential $\mu(\rho)$ and the individual energies
$\epsilon(\kv,\rho)$, we have: 
\be
\mathrm{n}(\kv,\rho)=g\left[1+{\rm e}^{-\beta X(\kv,\rho)}\right]^{-1}
&\;\;\; ; \;\;\;&
X(\kv,\rho)=\mu(\rho)-\epsilon(\kv,\rho)
\;,
\ee
with $\epsilon(\kv,\rho)$ the individual particle level of momentum $\kv$ 
and $g$ the level degeneracy.
The residual interaction term now reads:
\be
2 \frac{\delta \mathrm{E}^{\rm res}}{\delta \mathrm{N}^2}
&=&
\frac{1}{(2\pi)^6}
\int{{\rm d}\kv_1{\rm d}\kv_2
\mathrm{V}^{\rm res}_{\kv_1,\kv_2;\kv_1,\kv_2} \;
\frac{\partial X(\kv_1,\rho)}{\partial \rho} \frac{\partial \mathrm{n}(X(\kv_1,\rho))}{\partial X} \;
\frac{\partial X(\kv_2,\rho)}{\partial \rho} \frac{\partial \mathrm{n}(X(\kv_2,\rho))}{\partial X} }
\ee

For non-relativistic nucleons, with a Skyrme interaction, we have:
\be
X(\kv,\rho)=X(k^2,\rho)&=&\mu(\rho)-\left[\mathrm{U}(\rho)+\frac{k^2}{2\mathrm{m}^*(\rho)}\right]
=\nu(\rho)-\frac{k^2}{2\mathrm{m}^*(\rho)}
\;.
\ee
At zero temperature, the density derivative of the occupation number is a Dirac function:
\be
\mathrm{n}(k^2,\rho)&=&g\Theta(X(k^2,\rho))\\
\frac{\partial \mathrm{n}(k^2,\rho)}{\partial \rho}&=&g\frac{\partial X(k^2,\rho)}{\partial \rho}\delta(X(k^2,\rho))
\;.
\ee
Since $X=0$ for $k^2=k_{\rm F}^2$, the integral becomes
\be
2 \frac{\delta \mathrm{E}^{\rm res}}{\delta \mathrm{N}^2}
&=&
\left[ \frac{g k_{\rm F} {\rm m}^*}{(2\pi)^2}\frac{\partial X(k_{\rm F}^2,\rho)}{\partial \rho} \right]^2
\int {{\rm d}\cos\theta_1 {\rm d}\cos\theta_2 \mathrm{V}^{\rm res}_{\rm F}(\cos\theta_1,\cos\theta_2)}
\;.
\ee
Neglecting the angle dependence of the residual interaction,
we have $\mathrm{V}^{\rm res}_{\rm F}(\cos\theta_1,\cos\theta_2)=\mathrm{V}^{\rm res}_{\rm F}$.
Using the relation
\be
\frac{\partial X(\rho,k_{\rm F}^2)}{\partial \rho}&=&\frac{2\pi^2}{\mathrm{m}^*gk_{\rm F}}
\ee
we finally obtain that
\be
2 \frac{\delta \mathrm{E}^{\rm res}}{\delta \mathrm{N}^2}&=&\mathrm{V}^{\rm res}_{\rm F}
\;.
\ee

In order to generalize this result at finite temperature, 
let us now assume that the residual interaction elements 
$\mathrm{V}^{\rm res}_{\kv_1,\kv_2;\kv_1,\kv_2}$
are equal to a constant $\mathrm{V}^{\rm res}$.
Starting again from Eq.~(\ref{EQ-an:dEres-integ}) we simply have:
\be
2\frac{\delta \mathrm{E}^{\rm res}}{\delta \mathrm{N}^2}
&=&\mathrm{V}^{\rm res}\frac{1}{(2\pi)^6}\int{{\rm d}\kv_1 {\rm
    d}\kv_2 \frac{\partial \mathrm{n}(\kv_1,\rho)}{\partial \rho}
  \frac{\partial \mathrm{n}(\kv_2,\rho)}{\partial\rho}} 
=\mathrm{V}^{\rm res}\left[\frac{\partial}{\partial \rho}
  \int{\frac{{\rm d}\kv}{(2\pi)^3} \mathrm{n}(\kv,\rho)} \right]^2 
=\mathrm{V}^{\rm res}
\ee
whatever the form of $\mathrm{n}(\kv,\rho)$, since 
$\rho=\int{\frac{{\rm d}\kv}{(2\pi)^3} \mathrm{n}(\kv,\rho)}$.


\section*{Acknoledgments}
C.D. and J.M. want to thank N. Van Giai for interesting discussions 
concerning the RPA framework in the unstable phase diagram region. 
C.D. thanks C. Provid\^encia for fruitful discussions on the Vlasov formalism.
This work has received the support of the project ``mod\'elisation des
\'etoiles \`a neutrons'' supported by CNRS-IN2P3.


\end{document}